\documentclass[12pt,preprint]{aastex}
\begin{document}
\title{Saving Planetary Systems: Dead Zones \& Planetary Migration}
\author{Soko Matsumura\altaffilmark{1},  Ralph E. Pudritz\altaffilmark{2}}
\affil{Department of Physics \& Astronomy, McMaster University,
Hamilton, ON L8S 4M1, Canada}
\email{soko@physics.mcmaster.ca, pudritz@physics.mcmaster.ca}
\and
\author{Edward W. Thommes}
\affil{Canadian Institute for Theoretical Astrophysics, University of
Toronto, 60 St. George Street, Toronto, ON M5S 3H8, Canada}
\email{thommes@cita.utoronto.ca}

\altaffiltext{1}{Current location: Department of Physics \& Astronomy, 
Northwestern University, 2145 Sheridan Road, Evanston, IL 60208-0834, USA, e-mail: soko@northwestern.edu}
\altaffiltext{2}{Origins Institute, ABB 241, McMaster University, 1280
Main Street West, Hamilton, ON, L8S 4M1, Canada}
%
%
%--- Abstract ---
\begin{abstract}
The tidal interaction between a disk and a planet leads to the
planet's migration. It is widely believed that this mechanism
explains the variety of orbital radii of extrasolar planets. A
long-standing question regarding this mechanism is how to stop the
migration before planets plunge into their central stars.

In this paper, we propose a new, simple mechanism to significantly
slow down planet migration, and test the possibility by using a
hybrid numerical integrator to simulate the disk-planet interaction.
The key component of the scenario is the role of low viscosity regions in 
protostellar disks known as dead zones. A region of low viscosity affects 
planetary migration in two ways. First of all, it allows a
smaller-mass planet to open a gap, and hence switch the faster type
I migration (pre-gap-opening migration) to the slower type II
migration (post-gap-opening migration). Secondly, a low viscosity
slows down type II migration itself, because type II migration is
directly proportional to the viscosity.

We present numerical simulations of planetary migration in disks by using a 
hybrid symplectic integrator-gas dynamics code. Assuming that the disk 
viscosity parameter inside the dead zone is \(\alpha=10^{-4}-10^{-5}\), 
we find that, when a low-mass planet (e.g. 1 - 10 Earth masses)
migrates from outside the dead zone, its migration is stopped due
to the mass accumulation inside the dead zone. When a low-mass
planet migrates from inside the dead zone, it opens a gap and slows
down its migration. A massive planet like Jupiter, on the other
hand, opens a gap and slows down inside the dead zone, independent
of its initial orbital radius.  The final orbital radius of a Jupiter mass planet depends 
on the dead zone's viscosity.  For the range of \(\alpha\)'s noted above, this can vary 
anywhere from 7 AU, to an orbital radius of 0.1 AU that is characteristic of the hot 
Jupiters. 
%
%from the Jupiter like orbital radius (\(\sim 7\) AU) to the hot Jupiter like radius 
%(\(\sim 0.1\) AU).  
%Finally, we show that there are 
%characteristic values for the disk viscosity parameter in the dead zone 
%(\(\alpha\sim 10^{-4}\)) that can account for the close-in orbits (\(\sim 0.1\) AU) of 
%massive planets.
%
\end{abstract}
%
%--- Introduction, Section 1 ----------------------------------------------------------------------
\section{Introduction}
Since the first discovery of an extrasolar planet around a
solar-type star in 1995 \citep{Mayor95}, surveys for extrasolar
planets have sampled several thousand solar-type stars. Out of
these, at least \(5-25\%\) harbor Jupiter mass planets within \(\sim
5\) AU of their central stars \citep[e.g.][]{Lineweaver03,Butler06}.
%To date, more than 180 extrasolar planets have been discovered.
Most extrasolar planets are of Jupiter mass \(M_{J}\), or even larger (up to
\(\sim 10 M_{J}\)), and tend to orbit very close to the central star
--- sometimes even closer than Mercury is to the Sun
\citep[e.g.][]{Udry03}. 
These close-in planets, which have orbital radii less than about 0.1 AU, are 
called ``hot Jupiters" \citep[e.g.][]{Mayor95,Marcy96,Santos05}.
Arguably the best explanation that we have
for these planets is that they were formed further out in the disks,
and then migrated to the current locations
\citep[e.g.][]{Goldreich80,Lin93,Ward97,Masset03,Artymowicz04}. This
raises several interesting questions about the evolution of
planetary systems. As an example, does the current estimate of the
percentage of stars harboring planets represent the overall
probability of a planetary system to survive, or does it rather
suggest that only in \(5-25\%\) of cases planetary systems undergo
enough migration to place Jupiter-like planets within \(5\) AU? In
this latter view, there would be more planets beyond \(5\) AU,
waiting to be detected in future programs.

The problem with migration as it is currently understood, is that
planet-disk interaction is too efficient. Planets rapidly migrate
within a gaseous disk and plunge into their central stars in less than a few 
million years for disk
models with standard values of the disk viscosity. This raises the question of 
whether or not planetary systems typically survive at all. In this work, we
will demonstrate that dead zones (low viscosity regions in disks)
play a crucial role in saving planetary systems by significantly
slowing planetary migration.

%Why do planets migrate in disks?
Planets migrate as they tidally interact and exchange angular
momentum with their disk at Lindblad or corotation resonances.
%The corotation torque is proportional to the surface mass density
%gradient \citep[e.g.][]{Goldreich79,Tanaka02}, and becomes
%negligible for disks with the surface mass density profile of
%\(\Sigma\propto r^{-3/2}\).
The different kinds of planet migration that arise are generally
divided into three types (type I, II, and III). Type I migration
occurs for planets that are not massive enough to open a gap in
protostellar disks (e.g. Earth mass and 10 Earth mass planets). Such
planets most strongly interact with the disk at Lindblad resonances,
and hence the effect of corotation torque is negligible
\citep[e.g.][]{Goldreich80}. Usually, planets feel a stronger outer
torque than an inner torque, and thus the total torque is negative
and the overall migration is inward, relative to the disk. The
migration speed is proportional to the planetary mass \(M_p\), disk
surface mass density \(\Sigma\), and inversely proportional to the
square of the disk aspect ratio \(h/r\), where \(h\) is the pressure
scale height and \(r\) is the disk radius
\citep[e.g.][]{Goldreich80,Lin93,Ward97}.
%
%As a planet accretes gas from the disk and increases its
%mass, the torque it exterts on the disk become stronger.
%The inner disks lose angular momentum and moves inward, while the
%outer disk gains angular momentum and moves outward, i.e. the planet
%opens a gap in the disk.
%The gaps, however, don't completely open for all planets, because the
%viscosity in differentially rotating disks forces the gap's inner edge
%material to flow outward, and the outer edge material to flow inward,
%i.e. the gap tries to close again.
%In other words, the gap-opening process is a fight between the
%disk's viscous torque and the planet's tidal torque.
%The gap completely opens only when the angular momentum transfer rate
%by the tidal torque exceeds that by the viscous torque.

Type II migration occurs for planets that are massive enough to open
a clear gap in the disks (roughly Jupiter mass and up). Such planets
are locked in the gap, and coupled to the disk through the Lindblad
resonances which are distant enough from the planets to fall beyond the gap edge.
%Such planets are tidally locked in the gap, because Lindblad resonances fall
%inside the gap and planets cannot exchange angular momentum with the
%surrounding disk.
They migrate as the gaseous disk accretes toward
the central star due to the disk's viscosity. As long as the disk mass is large compared to the planet mass, 
the migration speed is proportional to the disk's viscosity parameter \(\alpha\) and the
square of the disk's aspect ratio. Type II migration is generally
one to two orders of magnitude slower than type I migration
\citep{Ward97}.

Finally, type III migration pertains to planets with an intermediate
mass (e.g. Saturn mass planets), which are just heavy enough to
start opening a gap. These planets feel the effect of a sharp
transition of the surface mass density at the edges of gaps. Since
the corotation torque is proportional to the surface mass density
gradient \citep[e.g.][]{Goldreich79,Tanaka02}, its effect dominates
the exchange of angular momentum between a planet and a disk. The
angular momentum is transferred as the disk material which is
trapped in horseshoe orbits passes by the planet. This may lead to a
rapid inward or outward migration depending on the initial migration
direction of the planet
\citep[e.g.][]{Goldreich80,Ward92,Masset03,Artymowicz04}. The
migration speed is comparable to the type I migration
\citep{Masset03}.

For standard disk models, all types of migration are primarily
inward, and their timescale is at least one to two orders of
magnitude shorter than the disk's lifetime (\(10^6-10^7\) years).
Therefore, in order to avoid being swallowed by their central stars,
the planets' migration has to be significantly slowed down or
stopped somehow. Otherwise, giant planets would only have a chance
to form once the gas disk is already significantly depleted
\citep{Thommes06}.
%(Thommes \& Murray 2006, ApJ 644, 1214)

A number of scenarios have been proposed to solve this problem. Some
scenarios stop planet migration very close to the central stars
either by an inner magnetospheric cavity in the disk
\citep[e.g.][]{Shu94,Lin96}, or by the tidal interaction with the
star \citep[e.g.][]{Lin96}. Also, an overflow of the Roche lobe between the central star and a planet could reverse planet migration \citep{Trilling98}, because a planet
loses mass to the central star and therefore moves outward to
conserve the angular momentum of the system. However, these
mechanisms cannot explain the planets orbiting farther than the
immediate vicinity of the star.

Another scenario predicts that the planets stop migration when they
move from a non-magnetized region of a disk into a magnetized region
\citep{Terquem03}. This is because the outer torque due to the
magnetic resonances disappears in such a case, and because the
contribution from the magnetic resonances dominates that from the
Lindblad resonances. This mechanism, however, cannot stop type II planet migration.

The torque fluctuations due to the magnetorotational instability
(MRI) turbulence which were recently discovered in numerical
simulations \citep[e.g.][]{Laughlin04,Nelson04,Nelson05} may be able
to prolong the lifetime of some planets \citep{Johnson06b}.  This
mechanism, however, won't be able to explain the survival of planets
heavier than about 10 Earth masses.

We show in this paper that dead zones can stop or slow down planetary migration
very efficiently. The underlying assumption here is that the major source of disk's viscosity is the magnetorotational instability (MRI) turbulence \cite[e.g.][]{Balbus91}. A dead zone is a dense region in a disk which is
poorly ionized, and hence MRI inactive and nearly inviscid \citep[e.g.][]{Gammie96}.
%Since the MRI turbulence is the most promising source of viscosity
%\citep[e.g.][]{Balbus91}, inside the dead zone is expected to be
%nearly inviscid.
%Most work regarding planet formation and migration
%assumes that the viscosity in disks is spatially uniform
%\citep[e.g.][]{Pollack96,Hubickyj05,Alibert05}.
Although typical disks are expected to have dead zones
\citep[e.g][hereafter MP03, MP05, and MP06respectively]{Gammie96,Glassgold97,Sano00,Fromang02,Semenov04,Matsumura03,Matsumura05,Matsumura06},
their effect on the evolution of protostellar systems has received little
attention. Since the disk's viscosity affects the rate of planetary
migration, and since planet formation occurs concurrently with
migration, it is important to take account of such variations in
disk's viscosity.

In this paper, we study planet migration in disks with dead
zones by means of state of the art numerical simulations as well as
analytical calculations.
%These are designed to better follow the interaction of planets and disks.
%In this paper, we consider the planet migration and formation by taking account of
%the dead zone --- poorly ionized region in a protostellar disk which
%has no magneto-rotational instability (MRI) turbulence
%\citep{Gammie96}, and hence expected to be nearly inviscid.
We first introduce our disk models and numerical methods, and show
the disk evolution with a dead zone in \S \ref{chap2}. Then we consider the
case of planetary migration beginning from beyond the dead zone, and
compare it with the case of disks with no dead zone (\S \ref{chap3}). We then
present the planet migration from inside the dead zone, focusing on
a width of the gap opened by a planet in \S \ref{chap4}. The migration in
disks with various viscosity parameters is briefly discussed in \S \ref{chap5}. 
%We also
%take account of the dead zone evolution and mass
%accretion through the surface layers in \S \ref{chap6}.
%%We discuss the percentage ratio of observed planets and the total
%%possible number of planets in relation to the disk's parameters in
%%\S 7.
Finally, the discussion and conclusion are presented in \S \ref{chap7}.
%
%\begin{itemize}
%\item purposes of the paper
%\begin{itemize}
%\item comparing disks with and without dead zones
%\item considering a dead zone's effect on type I \& II migration
%\item dead zone evolution -- it's not totally dead
%\item planet scattering
%\end{itemize}
%\end{itemize}
%
%--- Section 2 -------------------------------------------------------------------------------------
\section{Disk Models \& Numerical Simulations} 
\label{chap2}
\subsection{Numerical methods}
We perform numerical simulations of planet migration in
protostellar disks using a hybrid numerical code, which combines an
N-body integrator with a simple disk model \citep{Thommes05}.
The N-body part is based on the symplectic integrator SyMBA
\citep{Duncan98}, which applies the method of \cite{Wisdom91}
with an adaptive timestep to resolve close encounters between massive
bodies.
The disk part of the code evolves viscously as well as through angular
momentum exchange with the embedded planets.

In this code, the disk is treated as one-dimensional in the radial
direction, so the disk properties are averaged vertically and
azimuthally. However, three-dimensional effects are implicitly
included in the code --- the vertical direction effect comes in
through the pressure scale height (or equivalently, the sound speed) 
%and the torque density due to the planet-disk interaction, 
and the azimuthal direction through the torque density, which represents the effect of azimuthally asymmetric 
structure, i.e. the spiral density waves launched at Lindblad resonances (see Appendix B).
%the surface mass density as well as the torque density (see Appendix B).

By combining the continuity and angular momentum equations, the
evolution of the surface mass density of a disk is calculated as
follows \citep[e.g.][]{Pringle81}:
\begin{equation}
%\frac{\partial \Sigma}{\partial t} &=& \frac{1}{r}\frac{\partial}{\partial r}
%\left[3r^{1/2}\frac{\partial}{\partial r}\left(\nu \Sigma
%r^{1/2}\right) - \frac{r^{1/2}}{\pi \sqrt{GM_{*}}}\frac{\partial
%T}{\partial r} \right] \nonumber \\
%
\frac{\partial \Sigma}{\partial t} =
\frac{1}{2\pi r}\frac{\partial}{\partial r}
\left[ \left[\frac{\partial}{\partial r}\left(\Omega r^2\right)\right]^{-1}
 \frac{\partial}{\partial r}\left(T_{\rm viscous}-T_{\rm tidal}\right) \right] \label{surfevol} \ ,
\end{equation}
where \(T_{\rm viscous} = 3\pi r^2 \nu \Sigma_{gas} \Omega\) is the
viscous torque (\(\nu=\alpha c_s h\) is the viscosity,
\(\Sigma_{gas}\) is the gas surface mass density, and \(\Omega\) is
the orbital frequency), and \(T_{\rm tidal}\) is the tidal torque
(see Appendix B for an expression). Since planetary migration is
mainly affected by the principal Lindblad resonances
\citep[e.g.][]{Goldreich80,Ward97}, we neglect higher order
resonances as well as corotation resonances. We assume a nearly
Keplerian disk \(\kappa\sim\Omega\), and therefore calculate the
tidal torque between the 2:1 and 1:2 resonances.  Planetary
eccentricities are assumed to be nearly zero throughout this paper.
%This assumption is valid as long as the planet's orbit is nearly
%Keplerian \citep[e.g.][]{Goldreich80}.

Each time step of the above equation is solved in three steps.
First, the disk is evolved due to only the viscous diffusion (e.g.
under the influence of the first term in Eq. \ref{vtypeI} alone) for half a time step.
Then it is evolved due to the planet-disk interaction (e.g. under
the influence of the second term) for a full time step. Finally, the
disk is evolved for another half a time step under the influence of
the viscous diffusion. The planet's orbit is calculated three
dimensionally in a similar manner.
%{\bf In all cases planetary eccentricities are zero.}

\subsection{Initial Conditions}
We set up the initial disks based on our previous work (MP06).
We consider both disks with and without dead zones.
For the former case, we use the dead zone radii which were obtained in our
previous work, and assume the lower viscosity parameter inside the
dead zones.

Since the detailed explanation of our disk models and the way we
determined the size of the dead zones are documented elsewhere (MP03, MP05,
and MP06), we simply summarize our previous results here.
%The existence of dead zones are extensively studied by many authors
%including MP03, MP05a, and MP05b.

MP06 showed that a typical dead zone extends from 0.04 to 13 AU in
protostellar disks. Since the size of the dead zones depend on both
the thermal and ionization structure of disks, we used the disk
models which reproduce the observed spectral energy distributions
(SEDs) very well \citep{Chiang01,Robberto02}, and calculated the
electron density within them (see MP03, MP05 as well). We took
account of the ionization by X-rays from the central stars, cosmic
rays, radioactive elements, and thermal collisions of alkali ions,
as well as the recombination by molecular ions, metals, and
grains. The innermost disk (less than about 0.04 AU) is magnetically
active since the collisional ionization is efficient, and since
grains, which are an efficient recombination source, are evaporated
due to the high temperature there. The outer part of the disks
(larger than about 13 AU) are magnetically active since the disks
are less dense, and therefore well-ionized. Similar work done by
\cite{Semenov04} gives roughly the same results. For simplicity, we
assume that the disk within \(13\) AU is dead in this paper (see
also \S 6).  
%This assumption does not affect the results, since the
%numerical resolution of our code is not very good within \(1\) AU.

Throughout this paper, we use a fiducial value for the viscosity
parameter \(\alpha=10^{-5}\) for inside the dead zones, and
\(\alpha=10^{-2}\) for outside it (active zones). These values agree
well with the three-dimensional numerical MHD simulations of the
layered accretion disks done by \cite{Fleming03}. We note however
that real disks may have a significant range for the viscosity
parameter centered on these fiducial values (see \S 5). 
%Finally, the
%low value for dead zone viscosity does not prevent accretion onto
%the central star. This latter accretion flow proceeds through the
%magnetically well coupled, and hence higher viscosity, surface layer
%of the disk. We include such an effect in \S 5.7.

The initial disks have a gas surface mass density structure of
\(\Sigma_{gas}=\Sigma_{0,gas}(r/AU)^{-3/2}\) where
\(\Sigma_{0,gas}=10^3 \ {\rm g \ cm^{-2}}\), and the temperature
structure of disk models by \cite{Robberto02}. We assumed the solid
surface mass density of
\(\Sigma_{solid}=\Sigma_{0,solid}(r/AU)^{-2}\) where
\(\Sigma_{0,solid}=270 \ {\rm g \ cm^{-2}}\), following
\cite{Pollack96} and \cite{Alibert05}. Although the solid materials
or planetesimals don't directly affect the results of the current
paper, they will become important during the formation of planets in
disks with dead zones. We will discuss this in a subsequent paper.

We perform simulations on only low-mass planets (1 and 10 Earth mass
planets) and massive planets (1 Jupiter mass planet). We don't take
account of type III migration, which is shown to be effective for
intermediate mass planets like Saturn
\citep[e.g.][]{Goldreich80,Ward92,Masset03,Artymowicz04}. This is
because there has been no reliable way as yet to evaluate the
corotation torque in an azimuthally averaged way.

\subsection{The dead zone's effects on disk surface mass density}
We first show how a dead zone with a fixed radius would change the
disk's surface mass density profile.  
Here, we consider the disk with no planets.  We assume that the initial surface mass density
has a smooth power law distribution (\(\Sigma \propto r^{-3/2}\),
see the last subsection), and that the viscosity parameter is
\(\alpha=10^{-5}\) inside the dead zone radius \(r_{DZ}=13 \ {\rm
AU}\), and \(\alpha=10^{-2}\) outside it.

Our main assumptions are (1) the dead zone has a fixed outer radius 
(typically, \(r_{DZ}=13\) AU), (2) the temperature profile does not change
throughout the simulation, and (3) there is no surface layer
accretion. Naturally, each of these factors affects disk evolution
significantly. We will discuss this issue in a subsequent paper.  
%We show in \S 5.7 that disks still evolve in a similar
%manner even when we include the above effects.
%do develop a density jump even when we include the above effects.
Here, we merely consider a simplified model, which captures a trend
of the evolution of such a disk, and discuss the effects of a dead
zone on planet migration.
%density jump on planet migration.

Fig. \ref{fig1} shows the evolution of the disk's surface mass density.  
Since the
mass accretion speed is proportional to the viscosity parameter
\(\alpha\), the disk outside a dead zone accretes three orders of
magnitude faster than the dead zone. Therefore, the disk quickly
develops a steep surface mass density gradient at this radius
(within \(t\sim10^4\) years). The mass accreting from outside the
dead zone accumulates at the edge of the dead zone, and slowly
spreads over the dead zone. This mass accumulation makes a jump in
the surface mass density profile, which eventually reaches a three
to four orders of magnitude difference between dead and active zones
\footnote{It should be pointed out that such a steep density jump 
makes a disk locally Rayleigh unstable (\(\kappa^2<0\), see Appendix B for
an expression).  This is likely to reduce the gradient of surface
mass density somewhat. We will leave this as a future work.}

When the surface mass density increases as seen in the simulation, the
disk becomes optically thick, and therefore the ionization rates decrease.  
This ensures that a dead zone is even more ``dead", because the MRI 
turbulence requires a disk to be
well-ionized. We can estimate dead zone's effects on planet
migration as follows. Assuming a vertical hydrostatic equilibrium
for an ideal gas, we can write the disk aspect ratio as
\(h/r=\sqrt{(T/T_c)(r/R_*)}\) where \(T_c=k_BR_*/(GM_*\mu m_H)\)
(\(k_B\) is Boltzmann constant, \(G\) is the gravitational constant,
\(\mu\) is the mean molecular weight, \(m_H\) is the hydrogen mass,
and \(R_*\) and \(M_*\) are stellar radius and mass respectively).
With this, the type I and type II migration speeds can be written as
\citep[e.g.][]{Ward97}:
\begin{eqnarray}
v_{\rm type I} &\propto& -2 \frac{M_p}{M_*}(r_p\Omega_p) \frac{\pi
\Sigma_p r_p^2}{M_*} \left(\frac{T_p}{T_c}\frac{r_p}{R_*}\right)^{-1} \label{vtypeI} \\
v_{\rm type II} &=& -\frac{3}{2} \alpha (r_p\Omega_p)
\left(\frac{T_p}{T_c}\frac{r_p}{R_*}\right) \label{vtypeII} \ ,
\end{eqnarray}
where the subscript \(p\) indicates that the values are evaluated at
a planet's orbital radius \(r_p\). 
In typical protostellar disks, two order of magnitude increase in surface mass density 
leads to a factor of a few increase in temperature \citep[e.g.][]{Lecar06}, and hence 
the temperature evolution is negligible compared to the surface mass density evolution.  
Since the surface mass density increases inside the dead zone, the type I migration 
speed there is expected to be {\it faster} than that outside the dead zone. 
On the other hand, the type II migration
speed is expected to be {\it slower}, because the disk's viscosity
parameter \(\alpha\) is smaller.
Therefore, we expect that at least type II migration is slowed
inside a dead zone.

Another important effect of a dead zone on planet migration is that
the gap-opening mass becomes very small inside a dead zone.  Fig.
\ref{fig2} shows the gap-opening mass in our initial disk model with
a dead zone.  The upper dashed line shows the gap-opening mass
expected for a viscous disk with \(\alpha=10^{-2}\), while the lower
dashed line shows that for an inviscid disk.  A solid line is the
expected gap-opening mass throughout the disk.  Outside the dead
zone, planets open gaps when the tidal torque exceeds the viscous
torque \citep[e.g.][]{Lin93,Bryden99}:
\begin{equation}
\frac{M_p}{M_*}\gtrsim \sqrt{40\alpha\left(\frac{h_p}{r_p}\right)^5}
\label{gomass} \ ,
\end{equation}
which represents the upper dashed line. Outside the dead zone, this
corresponds to Jupiter or larger mass planets.  On the other hand, these 
results show that even a
terrestrial mass planet can open a gap inside the dead zone, since
the viscosity there is very small. The gap-opening mass in a
gravitationally stable, low viscosity disk (\(\alpha\sim10^{-5}\))
is well approximated by that in an inviscid disk with a reasonable
disk aspect ratio \(h/r \gtrsim 0.02\) \citep{Rafikov02}:
\begin{equation}
\frac{M_p}{M_*}\gtrsim \frac{2}{3}\left(\frac{h_p}{r_p}\right)^3
\min\left[5.2 Q^{-5/7}, \
3.8\left(Q\frac{r_p}{h_p}\right)^{-5/13}\right] \label{gomass_dead}
\ ,
\end{equation}
which represents the lower dashed line.  Here, we see the
gap-opening mass reaches an Earth mass at \(\sim 0.7\) AU.
%In an inviscid disk, the
%gap-opening mass is about an Earth mass at \(\sim 0.7\) AU, which
%corresponds to the same gap-opening mass in a viscous disk with
%\(\alpha=i10^{-4}-10^{-5}\).
Therefore, we expect that a planet with \(M_E\) or \(10 M_E\)
opens a gap inside a dead zone, and hence migrates at a slower type
II rate afterwards.

These terrestrial, gap-opening planets may execute type III
migration instead of type II migration inside the dead zone. In such
a case, these planets may migrate rapidly, at about one-third of the
type I migration rate \citep{Masset03}. If it were the case, the
planets would be trapped at the inner edge of the dead zone due to
the positive corotation torque \citep{Masset06}. In our dead zone
with \(\alpha\sim10^{-5}\), type III migration is unlikely because
the corotation torque tends to saturate in such a low viscosity
disk.  We show this analytically in Appendix A.

When planets are migrating from outside the dead zone, the situation
is more complicated.  The density jump affects both Lindblad and
corotation torques.  For the former case, the inner torque becomes
larger, because the torque is proportional to the surface mass
density.  If the density increase is large enough, the inner torque
exceeds the outer torque in magnitude, and the differential torque
becomes positive --- i.e. planet migration is directed {\it
outward}. We present a detailed analysis of this result in
Appendix B. On the other hand, for the corotation torque, such a
sharp surface mass density gradient would increase the amount of
material that gets caught in the inner horseshoe orbit, and hence
lead to the rapid {\it inward} migration. 
Therefore, we need to evaluate both Lindblad and corotation torques to
determine the exact direction of migration.  However, we will show
in Appendix B that the positive Lindblad torque starts acting far
away from the density jump.  Since the corotation torque has a
significant effect only at the density jump, it is likely that the
overall migration is directed outward due to the positive net
Lindblad torque.
%However, such a combined study is beyond the scope of this paper.
Therefore, we only take account of the effect of the Lindblad torque
in our calculation.
%We note, however, that when the
%migration is directed inward and the planet is drawn into the dead
%zone, it will be exactly like the migration inside the dead zone,
%which we will show in \S 4.

In the following sections, we follow planetary migration from an
initial orbit in which a planet is outside the dead zone (\S 3) as
well as inside the dead zone (\S 4).
%
%In the next section, we will analytically study the effect of such a
%density jump on planet migration from outside the dead zone.
%so that we can use it in our 1D disk model.
%This is because we are using a 1D disk where it is difficult to
%evaluate corotational resonances properly.
%When the planets are migrating inward, they lose their angular
%momentum as disk materials trapped in the horseshoe streamlines pass
%by them, and as a result execute a rapid migration.
%
%\begin{itemize}
%\item theory and simulations roughly agree with each other
%\begin{itemize}
%\item gap-opening masses are checked against our analytical work
%(Matsumura \& Pudritz, 2005)
%\item gap width is checked against Goldreich and Sari (2003)
%\end{itemize}
%\item plots gap-opening masses from simulations over the analytical
%gap-opening masses (add some data to Fig. \ref{fig1})?
%\end{itemize}
%
%--- Section 3 ------------------------------------------------------------------------------------
\section{Planet migration from outside the dead zones}
\label{chap3}
Here, we will focus on the migration of planets whose initial orbits
are outside the dead zone. Since a dead zone evolves much slower
than the disk annuli outside it, there should be a mass accumulation
inside the dead zone as seen in the last section.
%{\bf \color{red} I think it's better to say "inside" than "at the edge of" since the accumulation eventually spreads through the whole zone.  A density spike at the edge only exists while the disk readjusts itself from the initial smooth density profile when the simulation is started.  In Fig. 17 for example, you can see how the initial spike turns into just a step as the live zone drains into the dead zone.}
The resulting jump in surface density at the zone's outer edge can
stop the migration of the planet.
%if the Lindblad torque is stronger than the corotation torque. Here,
%Although we only take account of the Lindblad torque, it is unlikely
%that the corotation torque plays a significant role in this case as
%shown in Appendix B.
%and study its effect on planet migration. The
%interested readers should refer to Appendix B for an analytical
%study.
%In the following subsection, we analytically show this possibility, and then present the
%corresponding numerical results.
%
\subsection{Hybrid numerical simulations of planetary migration from outside
the dead zone}
%
%We now compare the planetary migration toward a density jump
%created at the outer edge of the dead zone for the different mass planets,
%namely planets with 1 Earth mass, 10 Earth masses, and 1 Jupiter mass.
We compute planetary migration towards the dead zone for
different mass planets, namely those with 1 Earth mass, 10 Earth
masses, and 1 Jupiter mass.  We choose their initial orbital radius
to be 20 AU, so that we can ensure that a density jump fully develops before planets approach 
the effective range of a surface mass density jump, and that even a
Jupiter mass planet starts as a type I migrator. Since massive planets probably form 
much closer-in, the migration rates we compute are upper limits. 

The former point is justified by Appendix B.  The results in Fig.
\ref{fig13} and \ref{fig14} suggest that planets start feeling the
effect of the mass accumulation from radii \(13 \ {\rm AU} \ <
r_p < 18.5 \ {\rm AU}\) (\(0.7<r_{\rm edge}/r_{p}<1\) with \(r_{\rm
edge}=13\) AU), depending on the steepness and magnitude of a
density jump as well as the disk thickness.

The latter point can be easily seen from Fig. \ref{fig2}.  For a
disk with \(\alpha=10^{-2}\), the gap-opening mass at 20 AU is
larger than \(1 M_J\) (see the upper line).  Therefore, all planets
start as type I migrators.
%
%Therefore, in order to be well outside the immediate effects of this
%density jump, we placed planets initially at 20 AU, in the MRI
%active, higher viscosity (\(\alpha=10^{-2}\)) region.
%
%We also choose our planetary masses at \(20\) AU to be beneath the
%threshold for opening gaps, so that we can distinguish type I
%migration from type II. Our choices are dictated by the gap-opening
%masses plotted in Fig. \ref{fig3}, wherein all chosen planets are
%below the gap-opening mass at 20 AU. Therefore, all of them are
%expected to start as a type I migrator. A Jupiter mass planet will
%probably begin to open a gap soon after the simulation starts, as is
%expected by its gap-opening radius (\(\sim 15\) AU in Fig.
%\ref{fig3}). On the other hand, planets of 1 Earth mass or 10 Earth
%masses will not open a gap in a standard viscous disk with a uniform
%parameter of \(\alpha=10^{-2}\) (see the upper dashed line in Fig.
%\ref{fig3} as well). If the disks have dead zones, however, 1 Earth
%mass and 10 Earth mass planets are expected to open gaps around
%\(1\) and \(5.5\) AU respectively. In our simulations, the orbit
%within \(1\) AU is not well-resolved, so it is unlikely to see the
%gap-opening by 1 Earth mass planet, but it is possible to observe
%the gap-opening by 10 Earth mass planet.

We first show the case for an Earth mass planet in Fig. \ref{fig3}.
%The left panel of Fig. \ref{fig4} shows the time evolution of a 1D disk with
%no dead zone (i.e. the alpha value is 0.01 everywhere) as well as an Earth mass
%planet migration in such a disk.
The left panel of Fig. \ref{fig3} shows the migration of a 1 Earth 
mass planet in a disk {\it without} a dead zone (i.e. the viscosity
has the parameter \(\alpha=10^{-2}\) everywhere). The y-axis shows
the radius of the disk, and the x-axis shows its time evolution.
Also shown are the contours of the disk's surface mass density. By
examining these contours, we can tell that the disk is accreting
toward the central star (y=0) as time goes on (toward right on the
x-axis). The thick black line is semi-major axis of the orbit of the
planet, which starts migrating from \(20\) AU. It is apparent that
the planet plunges into the star within about \(5\times10^6\) years (note that the migration rate 
decreases with semimajor axis).
This migration time roughly agrees with the one expected from the
theory \citep[\(\sim 2\times10^6\) years, see e.g.][]{Ward97}; the
softening parameter, which mimics the reduction of a torque with
vertical height (\(B=0.2\), see Appendix B as well), contributes to
making it longer.

The right panel of Fig. \ref{fig3} is the analogous figure for a
disk {\it with} a dead zone (i.e. the viscosity parameter
\(\alpha=10^{-5}\) inside and \(10^{-2}\) outside the dead zone). In this run, 
an Earth mass planet starts migrating from \(20\) AU, just outside the
outer dead zone radius \(13\) AU. As expected from the previous
section, the inward migration is halted before the planet reaches
the outer edge of the dead zone, and then the planet starts
migrating outward. When the planet moves away from the surface mass
density jump, the inner and outer tidal torques balance, and the
planet migration stops at around \(\sim 19.8\) AU. 
%\(\sim 19.5\) AU.
%Even after the planet moves out of the realm of the reversed torque imbalance
%due to the surface mass density jump, it keeps moving outward until it
%is slowed down.
%After \(10^7\) years, the planet is located at \(\sim 24\) AU.
The simulations were checked until they converged, by using smaller
and smaller radial steps.
%{\color{red} Odd, I wonder why it migrates so far out?  That's $r_{\rm edge}/r_{\rm pl}=0.54$.  The inner 2:1 resonance is at $1/2^{2/3}=0.63$ and I thought the code uses that as the inner edge for computing the inner torques.  I may be remembering wrong though...}
%
%The left panel of Fig. \ref{fig5} shows the surface mass density
%profile at the start of the \(1 M_E\) simulation (light-colored
%line), and after \(10^4\) years (dark line), while the right panel
%shows the corresponding profiles for \(10^6\) (light-colored line)
%and \(10^7\) years (dark line).
%%{\color{red} It would be good to plot the surface density at later times too, when the spike in surface density developes into a "step".}

These results can be well explained by using the analytic study
presented in Appendix B.  There, we model the density jump formed in
our simulations with a simple functional form, and evaluate the net
Lindblad torque a planet would feel. The results are shown in Fig.
\ref{fig13} and \ref{fig14} for various density jumps and widths.

The surface mass density evolution for an Earth mass planet is
exactly the same as Fig. \ref{fig1}, since the planet is totally
embedded in a disk, and does not open a gap. From the left panel of
Fig. \ref{fig1}, we can see the density jump is roughly two orders
of magnitude at \(\sim 13\) AU, which spans about two pressure scale
heights (here, the disk aspect ratio is \(h/r\sim 0.05\)).

The rightmost panels of the middle row in Fig. \ref{fig13} and
\ref{fig14} correspond to our simulations (a density jump \(F=100\),
and a jump width \(\omega=2h\)). These figures suggest that such planets 
will feel a net positive torque (i.e. experience the
outward migration) at a radius larger than \(r_{\rm edge}/r_p =
0.8\) and \(0.7\) (or equivalently, less than \(16.2\) AU or
\(18.6\) AU) respectively. Since the planet reverses its migration
%at \(\sim 19\) AU in Fig. \ref{fig4},
at \(\sim 18.3\) AU in Fig. \ref{fig3}, these analytical estimates are
in good agreement with the numerical simulations.

Similarly, from the right panel, we see that the density jump is
even larger at later times of the simulation (\(10^6-10^7\) years)
--- by about three orders of magnitude. Therefore, the planet feels a
stronger torque then compared to the estimates in Fig. \ref{fig13}
and \ref{fig14}, and its orbit reaches the equilibrium at around
\(19.8\) AU, where the inner and outer torques balance. This is the reason why we 
find the planet first halting its inward migration, and subsequently actually 
migrating some distance outward. 
%
%{\color{red} I can redo that calculation with B=0.2 so that it matches up}
%Since we are using the softening parameter \(B\) of \(0.2\), our 10 Earth mass planet
%case is expected to fall in somewhere between these two values.

We next turn to the results for the migration of a \(10M_E\) planet.
The left panel of Fig. \ref{fig4} shows its migration in a disk {\it
without} a dead zone. Again, it is clear that the planet plunges
into the central star within \(1.5 \times 10^5\) years, which agrees
well with the theory (\(\sim 10^5\) years).

The right panel of Fig. \ref{fig4} shows the result of a similar
simulation for a disk {\it with} a dead zone. This run follows a ten Earth mass
planet that starts migrating from \(20\) AU. Again, the planet is
repelled by the mass accumulation at the edge of the dead zone and
starts migrating outward at \(\sim 16.5\) AU, which agrees well with
the above estimate that is based on the analytical study in Appendix
B. Before the 10 Earth mass planet gets repelled by the mass
increase inside the dead zone, it migrates closer to the dead zone
than the Earth mass planet does. This is because more massive
planets migrate faster and hence get closer to the central star
before the mass accumulation starts affecting their migration. 
%Note that the density jump is fully developed before planets start migrating outward.
Similar to the 1 Earth mass case, the planet stops migrating after
it gets far from the surface mass density jump, at around \(\sim
19.5\) AU.
%The final orbital radius of the planet is \(\sim 29\) AU, which is
%about the orbital radius of Neptune (its mass is roughly \(17
%M_{E}\)).
Again, this result was checked until it converges by using smaller
radial steps.
%{\color{red} Again, I'm not sure why it's going so far out again.  I think this needs a closer look...}

The left panel of Fig. \ref{fig5} shows the migration of a 1 \(M_{J}\) planet 
in disks {\it without} a dead zone. We can see that the
planet plunges into the star within \(5\times10^5\) years.
%{\color{red} That doesn't match up to what the figure shows; there, the planet never has a chance to open a gap and hits the inner edge in less than 1 Myr}
This migration time roughly corresponds to the one expected from the
theory (\(\sim 10^6\) years).

The right panel of Fig. \ref{fig5} is the similar figure for a disk
{\it with} a dead zone. Once again, we place a Jupiter mass planet in an 
initial orbit of \(20\) AU, just outside the outer dead zone radius (\(13\) AU).
Here, the planet migrates into the dead zone as it opens a gap, and
drastically slows down its migration by the time it gets to 7 AU.
The planet is not stopped by the surface density increase in the
dead zone because it is locked into a clean disk gap, and is thus
constrained to move along with the accretion flow of gas through the
inner disk.
%The planet is not repelled by the surface mass density accumulation
%because it opens a gap and hence diminishes the inner torque.
%
%
%\begin{itemize}
%\item dead zone reverses the type I migration (see Fig. \ref{fig2})
%\item try a bit heavier planet to see whether they open a gap after
%they enter the dead zone?
%\end{itemize}
%
%
%--- Section 4 -------------------------------------------------------------------------------------
\section{Planet migration from inside the dead zone}
\label{chap4}
Here, we will focus on the migration of planets whose initial orbits
are within the dead zone. Intuitively, type I migration of
planets inside the dead zones won't be affected by the existence of
dead zones, since the type I migration does not depend on the disk's
viscosity. However, as we have seen in \S 2.3, the gap-opening mass
becomes smaller as the disk's viscosity decreases.  Therefore, we
expect that when the planetary mass is as large as \(M_E\) or
\(10M_E\), a planet inside a dead zone does not migrate straight
into the central stars as it would in a disk without a dead zone.
Instead, the planet ought to open a gap and thus slow down its
migration significantly. This is partly because the fast type I
migration is switched to the slower type II migration, and partly
because the type II migration speed itself is slower inside the dead
zones due to the lower viscosity there.
%We will first show the analytical estimates of gap-width due to
%massive and low-mass planets, and then show the results of the
%migration.
%
%
\subsection{Hybrid numerical simulations of planetary migration inside
the dead zone}
Fig. \ref{fig6} shows the evolution of the same disk as the right
panel of Fig. \ref{fig3}, and a 1 Earth mass planet starts migrating
from \(11\) AU, just inside the outer dead zone radius (\(13\) AU).
Here, planet migration appears to be much faster than in a disk
with no dead zone (left panel of Fig. \ref{fig3}). This is not
surprising if we recall that the type I migration speed is
proportional to the surface mass density (see Eq. \ref{vtypeI}).
%(\(v_{\rm type I}\propto -(r_p\Omega_p)(\pi\Sigma_pr_p^2/M_*)(h_p/r_p)^{-2}\)).
Due to a faster mass accretion from an active outer disk, the
surface mass density inside the dead zone increases, which leads to
faster migration.
%and because the surface mass density of the inner region becomes higher in a
%disk with the dead zone than without it, which makes type I
%migration even faster.
%{\color{red} But the figure shows the planet migrating much faster than in Fig. 4!  
%The reason for that is simple though; the surface density is much higher with the dead zone 
%than without it, so that makes Type I migration much faster}
Also note that we cut the planet migration at \(\sim 1\) AU, due to
the lower numerical resolution there, and that the planet is
expected to open a gap within 1 AU from Fig. \ref{fig2}.

Fig. \ref{fig7} shows the evolution of the same disk as the right
panel of Fig. \ref{fig4}, and a 10 Earth mass planet starts
migrating from \(11\) AU. Here, we find that the planet migrates in
quickly, and opens a gap at around \(4\) AU.  Note that, as expected
from \cite{Rafikov02}, the center of the gap lags behind the migrating planet.
%the migrating planet carves the gap {\it behind} it.  
Although the effect of the density wave propagation suggested by him is not included in our case, the physics is 
essentially the same.  The density deficit that is formed in front of the planet is refilled by 
migration, but that behind the planet is not. 
%This is because the density deficit behind a planet,
%which is formed due to the propagating density waves generated by
%the planet-disk interaction, cannot be refilled by migration.
%The type I migration speed is faster than the left panel of Fig.
%\ref{fig6} for the same reason as an Earth mass planet. 
The type I migration speed is faster than the left panel of Fig. \ref{fig4}
(a 10 \(M_E\) planet migration in a disk with no dead zone) for the same
reason as an \(M_E\) planet; the surface mass density increases inside the dead zone, which leads to the faster planetary migration (see Eq. \ref{vtypeI}).
After a gap-opening, its migration speed drastically slows as expected. 
The gap-opening radius (\(\sim 4\) AU) agrees well with the theoretical
estimate read from Fig. \ref{fig2} (\(\sim 5\) AU). 
The gap-opening mass for such a low-mass planet can be estimated from the damping length of density waves 
\citep{Rafikov02}:
%
%Also, the gap
%width of a 10 Earth mass planet estimated from Eq. \ref{cond12} is
%about \(0.37\) AU, agreeing well with the simulation (\(w \sim 0.4\)
%AU).
%
%
\begin{equation}
w \sim l_{\rm damping}=1.4\left(\frac{2}{3}\right)^{7/5}
\left(\frac{M_p}{M_*}\right)^{-2/5} \left(\frac{h}{r}\right)^{11/5}r
\label{gapeq2} \ .
\end{equation}
The estimated gap-width of about \(0.37\) AU agrees well with the simulation (\(w \sim 0.4\)
AU), although it should be pointed out that the damping of density waves is not included in our simulation.

Fig. \ref{fig8} shows the evolution of the same disk as the right
panel of Fig. \ref{fig5}, and a 1 Jupiter mass planet starts
migrating from \(11\) AU. The planet migrates inward as it opens a
gap, and significantly slows its migration inside the dead zone
around \(7\) AU. This shows that the dead zone slows down the type
II migration significantly.
%{\color{red} Actually Fig. 13 is the best illustration of Type II slowing down in the dead zone, since for the cases where alpha=0.01 outside the dead zone, the planet never even gets a chance to open a gap until it enters the dead zone.  So in those cases we're not really comparing type II inside and outside the dead zone, we're comparing type I outside with type II inside}
%
%\subsection{Gap width and the viscosity parameter}
%
%In this subsection, we will compare the gap width and the viscosity
%parameter with the ones expected from the theory.
%Recall that a gap opened by a massive planet is described as in Eq.
%\ref{gapeq}.
%Since the gap opens when the tidal torque exceeds the viscous torque, the gap width
%\(w\) can be estimated as follows
%\citep[e.g.][]{Goldreich03}:
For a massive planet like Jupiter, the excited density waves shock and damp immediately, so we can estimate the gap width by equating the tidal torque \(T_{\rm tidal} \sim C
\left(r/w\right)^3 \Sigma r^2 \left(r\Omega\right)^2 (M_p/M_{*})^2\)
(\(C=0.23\) in \cite{Lin93}) with the viscous torque \(T_{\rm viscous}\): 
\begin{equation}
\frac{w}{r} \sim \left(\frac{3\pi\alpha}{C}\right)^{-1/3}
\left(\frac{r}{h}\frac{M_p}{M_{*}}\right)^{2/3} \label{gapeq} \ .
\end{equation}
%\begin{eqnarray}
%T_{\rm viscous}&\sim&T_{\rm tidal} \nonumber \\
%3\pi\alpha\Sigma r^2 \left(\Omega h\right)^2 &\sim& C
%\left(\frac{r}{w}\right)^3 \Sigma r^2 \left(r\Omega\right)^2
%\left(\frac{M_p}{M_{*}}\right)^2 \nonumber \\
%\frac{w}{r} &\sim& \left(\frac{3\pi\alpha}{C}\right)^{-1/3}
%\left(\frac{r}{h}\frac{M_p}{M_{*}}\right)^{2/3} \label{gapeq} \ ,
%\end{eqnarray}
%
%where \(C\) is a constant which is \(0.23\) in \cite{Lin93}.

In Fig. \ref{fig9}, we plot the calculated gap width for a Jupiter
mass planet in a disk with \(\alpha=10^{-5}\) (dashed and dotted
lines, \(C=1\) and \(0.23\) in Eq. \ref{gapeq} respectively) as well
as the gap width determined from Fig. \ref{fig8} (solid line). The
planet starts migrating from \(11\) AU, and immediately opens a gap
whose size is comparable to the analytical estimates. For example,
from Fig. \ref{fig8}, we can see that 1 Jupiter mass planet inside
the dead zone with a viscosity alpha parameter \(\alpha=10^{-5}\)
forms a gap of width \(\sim 4\) AU, which is comparable to its orbital 
radius, while the analytically calculated gap size is \(\sim 3.5\)
AU at an orbital radius of \(4\) AU, which roughly agrees with the value
we got.
%
%--- Section 5 ------------------------------------------------------------------------------------
\section{Dependence upon Disk Viscosity Parameter: from Jupiter to hot Jupiters}
\label{chap5}
%In this subsection, we will discuss the effect of the viscosity parameter \(\alpha\).
Since there may be a significant range in values of the viscosity
parameter \(\alpha\) across the entire ensemble of observed protostellar
disks, we show a few other simulations
with different values of \(\alpha\) here.

In Fig. \ref{fig10}, we show the Jupiter mass planet migration (type
II migration) in a disk with \(\alpha=10^{-5}\) inside the dead zone
and \(\alpha=10^{-3}\) outside it. Similar to the right panel of
Fig. \ref{fig5}, the Jupiter mass planet migrates into the dead zone
as it opens a gap, and slows down its migration significantly. 
After \(10^7\) years, the orbital radius of the planet is \(\sim 7\) AU.
%{\color{red} actually it's already opened a gap before it gets to the dead zone; see comment above}

In Fig. \ref{fig11}, we show the Jupiter mass planet migration in a
disk with \(\alpha=10^{-4}\) inside the dead zone and
\(\alpha=10^{-2}\) outside it.
%Different from Fig. \ref{al0305} and the right panel of
%Fig. \ref{fig8},
It is difficult to see from the figure, but the planet starts
opening a gap {\it before} it enters the dead zone, at around 17 AU.  
The Jupiter mass planet migrates into the dead zone as it opens
a gap, and slows down its migration. Due to the higher viscosity
inside the dead zone, the migration is not slowed as much as Fig.
\ref{fig5} and \ref{fig10}.  
The final radius of the planet is
\(\sim 0.1\) AU.  Such planets may be good candidates for hot Jupiters.  In our fiducial
disk, the dead zone expands roughly between \(0.04\) and \(13\) AU.
If a planet migrates further in, it could be stopped by the inner edge of the dead zone
as suggested by \cite{Masset06}.

%but keeps on migrating to be swallowed by the central
%star within \(7\times10^6\) years.

In Fig. \ref{fig12}, we show the Jupiter mass planet migration in a
disk with \(\alpha=10^{-4}\) inside the dead zone and \(\alpha=10^{-3}\) outside it.
%Different from Fig. \ref{al0305} and the right panel of
%Fig. \ref{fig8},
Similar to the previous cases, the Jupiter mass planet migrates into
the dead zone as it opens a gap, and slows down its migration. The
migration inside the dead zone is faster than Fig. \ref{fig5} and
\ref{fig10}, but slower than Fig. \ref{fig11}. This is because the
planet opens a wider gap long before the planet in Fig. \ref{fig11}
does, and thus slows down earlier. After \(10^7\) years, the orbital radius of the planet is \(\sim 3.5\) AU. 
%This is because the viscosity outside the dead zone is weaker compared to Fig.
%\ref{fig14}, and hence the mass accumulation inside the dead zone is
%smaller, and the type I migration is slower.
%but keeps on migrating toward the central star.

These simulations indicate that planet migration is very sensitive to the disk's viscosity, 
and that dead zones with \(\alpha=10^{-4}-10^{-5}\) can place planets anywhere between 
\(\sim 0.1\) and \(\sim 7\) AU.   In other words, just by changing the viscosity parameter 
inside the dead zone by one order of magnitude we could get our Jupiter like planet 
(Fig. \ref{fig5}, \ref{fig10}, or \ref{fig12}) or a potential candidate for a hot Jupiter 
(Fig. \ref{fig11}). 
\section{Discussion and Conclusion}
\label{chap7}
We have studied planet migration in disks with dead zones, and compared
the results with the migration in disks with no dead zones.
Specifically, we studied the evolution of planets of mass \(M_{E}\),
\(10M_E\), and \(M_J\), which are initially placed in orbits both in
well coupled, magnetically active, and hence viscous regions as well as inside the dead
zone.

Throughout, we have only considered planets on circular orbits. In the case 
of type I migration, this is a reasonable assumption.  Planets are subject
to strong eccentricity damping, on an even much shorter timescale
than the orbital decay \citep[e.g.][]{Artymowicz93b,Papaloizou00},
%Papaloizou \& Larwood 2000,
and therefore core-sized bodies are likely to spend
almost all their time with very low eccentricities. Damping becomes
weaker once a gap opens (i.e. in type II regime).  In fact, the
disk-planet interaction for a gap-opening planet could lead to 
eccentricity enhancement \citep[e.g.][]{Artymowicz93b,Goldreich03},
but the issue of gas giant eccentricities is beyond the scope of
this work.

For planets migrating toward the dead zone in the active part of the
disk, there are some issues that we have not explicitly calculated. First
of all, as we have already mentioned, we have not included the
effect of the corotation torque.  This is potentially important when
planets are migrating toward the density jump generated by a dead
zone.  If the corotation torque is stronger than the Lindblad
torque, planets would be brought into the dead zone instead of being
repelled.  However, we showed that this is unlikely from the results of
Appendix B.  For steep density cases relevant to our simulations,
Fig. \ref{fig13} and \ref{fig14} show that the Lindblad torque changes
the sign from negative to positive at \(0.7\lesssim r_{\rm edge}/r_p
\lesssim 0.8\) (see panels with a density jump \(F=100\), and a jump
width \(\omega\sim h-2h\)). Therefore, the net Lindblad torque forces a planet to migrate {\it outward} at several pressure scale heights away from the dead zone edge.
This can be easily seen by noting \(5\lesssim
(r_p-r_{\rm edge})/h_{\rm edge} \lesssim 8.6\) with \(r_{\rm edge}=r_{DZ}=13 \
{\rm AU}\) and \(h_{\rm edge}/r_{\rm edge}\sim 0.05\).
%\(16.2 \ {\rm AU} \lesssim r_p \lesssim 18.6 \ {\rm AU}\)
%with \(r_{\rm edge}=r_{DZ}=13 \ {\rm AU}\) (see panels with a
%density jump \(F=100\), and a jump width \(\omega\sim h-2h\)).
Since the corotation torque has a significant effect on a length
scale of the density jump width, our results ensure that the
Lindblad torque starts acting on a planet {\it before} the
corotation torque becomes effective. Even if the corotation torque
becomes comparable to the Lindblad torque, it is unlikely the former
exceeds the latter. This is because the magnitude of the net
Lindblad torque close to the density jump (positive torque) is much
larger than that of the original Lindblad torque far from the jump
(negative torque).
%If, by any chance, the corotation torque exceeds the Lindblad torque and a planet enters a dead zone,
%the planet migration afterwards will follow the trend we have shown
%in \S 4 --- a type I migrator opens a gap, and slows its migration.

Secondly, turbulent fluctuations in active disk may cause
corresponding fluctuations in the torque felt by a planet during
type I migration \citep{Laughlin04,Nelson04,Nelson05,Johnson06b}.
%(Laughlin \& Steinacker \& Adams 2004, Nelson \& Papaloizou 2004).
In principal, if such fluctuations are large enough, this may make
it possible for planets/cores to ``jump the barrier" provided by the
edge of a dead zone, instead of being repelled by it. Such behavior
can be expected if the density fluctuations, $\delta \Sigma/\Sigma$,
become comparable to the density jump between active and dead zone.
When this happens, the planets would migrate just like the ones
inside the dead zone, i.e. type I migrators open a gap, and slow
their migration as seen in \S 4.
%Note, however, that when planets are brought into the dead zone,
%they will migrate just like planets inside the dead zone afterwards.
%We leave an investigation of this issue to future work.
%[NOT ADDRESSED: filling in of the gap from the active layers]  \\

Thirdly, in our simulations a density jump becomes steep enough so
that the disk is locally Rayleigh unstable (\(\kappa^2<0\), see
Appendix B for the expression).  This will smooth out the density contrast to 
some degree. Therefore, a density jump in a more
realistic disk would not be as steep as our models.  This could
affect the migration starting from outside the dead zone --- a
planet may not be reflected as strongly as we have shown.  
%In such a case, planet migration would be directed inward, which again
%migrates like planets inside the dead zone.  
We leave an investigation of this issue to future work.

In all three of these instances, we see that although the details of
the migration will change, the dead zone will still provide an
effective barrier through the onset of type II migration.

We list the conclusions of our paper below: \\

(i) Type II migration can be significantly slowed down inside the dead zones 
(see \S \ref{chap3} and \ref{chap4}). This is because the type II migration 
speed is directly
proportional to the disk's viscosity, and because the dead zones are
expected to have a very low viscosity. Jupiter mass planets open a
gap inside dead zones and almost stop their migration at
around \(4-8\) AU when the viscosity parameter is \(\alpha=10^{-5}\). \\
%
%(ii) For a migration of Jupiter mass planets to be effectively slowed down, the
%disk's viscosity has to be smaller than \(\alpha=10^{-4}\) (see \S 6).
%\\

(ii) Type I migrators are stopped at the edge of the dead zone (see
\S \ref{chap3}, as well as Fig. \ref{fig3} and \ref{fig4}). This is because
type I migration depends on the difference between inner and outer
torque strengths.  Due to the mass accumulation at the edge of the
dead zone, the magnitude of the inner torque becomes larger as a
planet approaches the edge, eventually canceling the outer torque.
This mechanism leaves both 1 Earth and 10 Earth mass planets at
\(\sim 20\) AU at early times, though they would move inward as the
dead zone contracts over time. \\

(iii) Type I migration inside the dead zone can be changed to a
slower type II migration (see \S \ref{chap4}, as well as Fig. \ref{fig7}).
Type I migration itself won't be slowed down by the presence of the
dead zones, since the migration speed does not depend on the disk's
viscosity. The gap-opening masses, however, become much smaller
inside the dead zones, and therefore even small mass planets (e.g.
\(M_E\) and \(10M_E\)) can open a gap there. 
Although we assume that density waves damp immediately, and hence do not 
include the wave damping effect proposed by \cite{Rafikov02}, our simulations show 
that a 10 Earth mass planet may open a gap at \(\sim 4\) AU, which slows down
its migration dramatically. \\

(iv) We also showed that the analytically calculated gap-width
agrees well with the results of simulations (see \S \ref{chap4}, as well as
Fig. \ref{fig9}). \\

(v) The dependence of migration on the viscosity parameter is
significant. If the viscosity outside the dead zone is measured with
\(\alpha\sim 10^{-2}-10^{-3}\), our simulations indicate that, after \(10^7\) years, the orbital radius 
of planets could vary from our Jupiter-like orbital radius \(\sim 3.5-7\) AU (see Fig. 
\ref{fig5}, \ref{fig10}, and \ref{fig12}) to hot Jupiter-like radius \(\sim 0.1\) AU 
(Fig. \ref{fig11}). \\
%the
%viscosity parameter inside the dead zone has to be less than
%\(\alpha=10^{-4}\) to efficiently slow down the type II migration
%there (see \S \ref{chap5}). \\

%(vi) Dead zones change their size significantly during the
%simulation time (\(10^6-10^7\) years) when we take account of the
%mass accretion through the surface layers. With viscosity parameters
%of \(10^{-2}\) outside the dead zone and \(10^{-5}\) inside it, the
%dead zone radius moves in from \(13\) AU to \(6\) AU within \(10^5\)
%years. \\
%
%(vii) If the distribution of disk parameters among different
%planetary systems is uniform, then the percentage of the observed
%planetary systems is well explained by the existence of low
%viscosity disks. If all disks have \(\alpha<10^{-4}\) inside
%\(10-20\) AU, and if planets start migrating from \(10-20\) AU, our
%calculation shows that \(20\%\) of planets will end up within \(5\)
%AU, which is the observation limit of the currently most powerful
%observing method. This percentage goes down to \(5\%\) for
%\(\alpha<4\times 10^{-5}\), while for \(\alpha=10^{-3}\), it goes up
%to \(30\%\). The range of values agrees well with the observational
%estimate of \(5-25\%\). Whether disks with dead zones are preferred
%or not will become apparent with future missions.  \\

Our approach so far has ignored the evolution of the dead zone which arises 
from the accretion of their surface layers onto the central star.  We also 
ignored the fact that planets accrete a
significant fraction of their mass while they migrate.
We will study planetary accretion during migration in a disk with an evolving 
dead zone in our next paper. \\
%One of the remaining questions is what happens to planet formation
%in a disk with a dead zone.
%We already developed the planet formation scheme in the disks with
%dead zones.
%In a subsequent paper, we will show these results.
%
%\begin{itemize}
%\item dead zone has important effects on the planetary evolution
%\item future work: gas-accretion
%\end{itemize}
%
%

We thank an anonymous referee for useful comments on our manuscript.
S. M. is supported by McMaster University, R. E. P. is supported by
a grant from the National Sciences and Engineering Research Council
of Canada (NSERC), and E. W. T. is supported by NSERC and CITA.
\appendix
\section{The saturation of corotation resonance inside a dead zone}
Here, we will discuss whether a gap-opening terrestrial mass planets
inside a dead zone executes type III migration instead of type II
migration.

The type III migration is driven by the corotation torque, which is
proportional to the inverse of vortensity gradient (\(d\ln
(\Sigma/B)/d\ln r\)) \cite[e.g.][]{Goldreich80}.  This gradient,
however, tends to be removed due to the libration of co-oribital
material, and therefore the corotation torque saturates, unless the
gradient is re-established by viscous torque
\citep[e.g.][]{Ward92,Masset01}. The corotation torque will not
saturate if the horseshoe libration timescale (\(t_{HS}\sim 4\pi
r_p/(1.5 \Omega_p x_s)\)) is longer than the viscous timescale
(\(t_{viscous}\sim x_s^2/(3\nu)\)) \citep{Masset03}:
\begin{equation}
\alpha > \frac{1}{8\pi}\left(\frac{M_p}{M_*}\right)^{3/2}
\left(\frac{h_p}{r_p}\right)^{-7/2} \ ,
\end{equation}
where the horseshoe zone half-width is taken to be \(x_s=0.96
r_p\sqrt{(M_pr_p)/(M_*h_p)}\), following \cite{Masset03}.

Since the gap-opening mass for low mass planets like \(M_E\) or \(10
M_E\) in a gravitationally stable, low viscosity disk
(\(\alpha\sim10^{-5}\)) is well approximated by Eq.
\ref{gomass_dead}, we replace the mass ratio in the above equation
with this, and obtain the following relation:
\begin{equation}
\alpha > 8.66\times 10^{-4} \left(\frac{h_p/r_p}{0.04}\right)
\left(\min\left[0.301 \left(\frac{Q}{54}\right)^{-5/7}, \
0.237\left(\frac{Q}{54}\frac{0.04}{h_p/r_p}\right)^{-5/13}\right]\right)^{3/2}
\ ,
\end{equation}
where \(Q=54\) and \(h/r=0.04\) are values estimated at 1 AU in our
fiducial disk. Here, the last term in the round bracket is of the
order of \(\sim 0.115\), and hence \(\alpha\gtrsim 10^{-4}\) is
needed to prevent the saturation of corotation torque. Therefore, it
is arguable whether the corotation torque plays a significant role
for planets migrating inside a dead zone, where the viscosity
parameter is very small (\(\alpha=10^{-4}-10^{-5}\)).
%
%---------------------------------------------------------------------------
\section{Type I migration toward a density jump --- the effect of Lindblad torque}
\begin{figure}
\plottwo{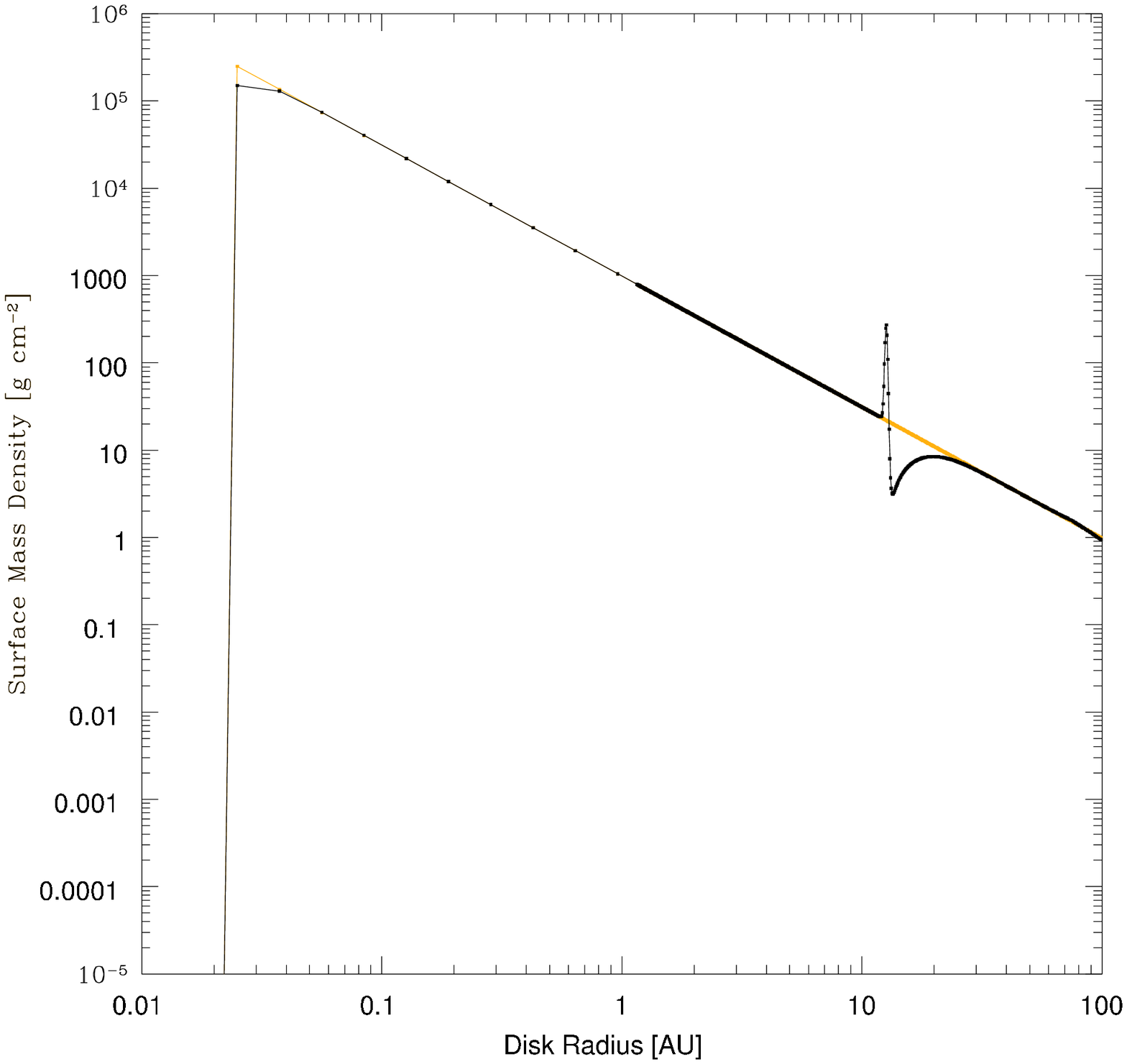}{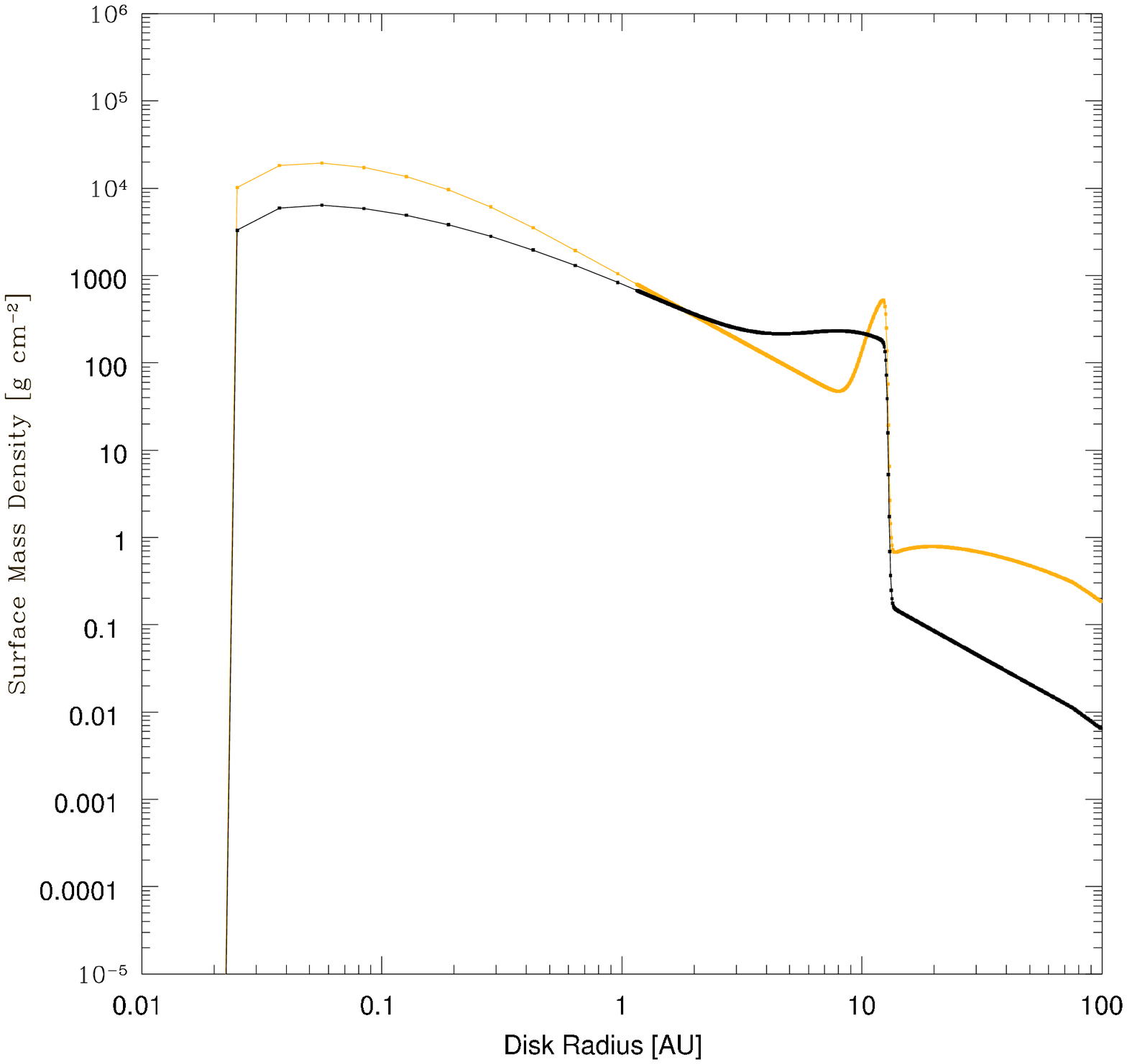}
%\unitlength1cm
%\begin{minipage}[t]{18.5cm}
%\hspace{0.5cm}
%\begin{minipage}[h]{16cm}
%\begin{picture}(5.5,6)
%\put(0,-1.5){\includegraphics[width=7cm]{fig5a.eps}}
%\end{picture}
%\hspace{2cm}
%\begin{picture}(5.5,6)
%\put(0,-1.5){\includegraphics[width=7cm]{fig5b.eps}}
%\end{picture}
%\end{minipage}
%\end{minipage}
%\begin{minipage}[t]{18cm}
%\vspace{0.5cm}
%
\caption[Surface mass density evolution of a disk with a dead
zone]{Left: The surface mass density evolution of a disk with a dead
zone. The dead zone radius is fixed at \(13\) AU. The light-colored
line is the initial power-law profile of the surface mass density,
and the dark line is the density profile after \(10^4\) years.
Right: The same for \(10^6\) years (light-colored line) and \(10^7\)
(dark line) years. The surface mass density is gradually flattened
due to the small viscosity inside the dead zone. \label{fig1}}
%\end{minipage}
\end{figure}

\begin{figure}
%\scalebox{0.3}{
%\includegraphics{fig3.eps}}
\plotone{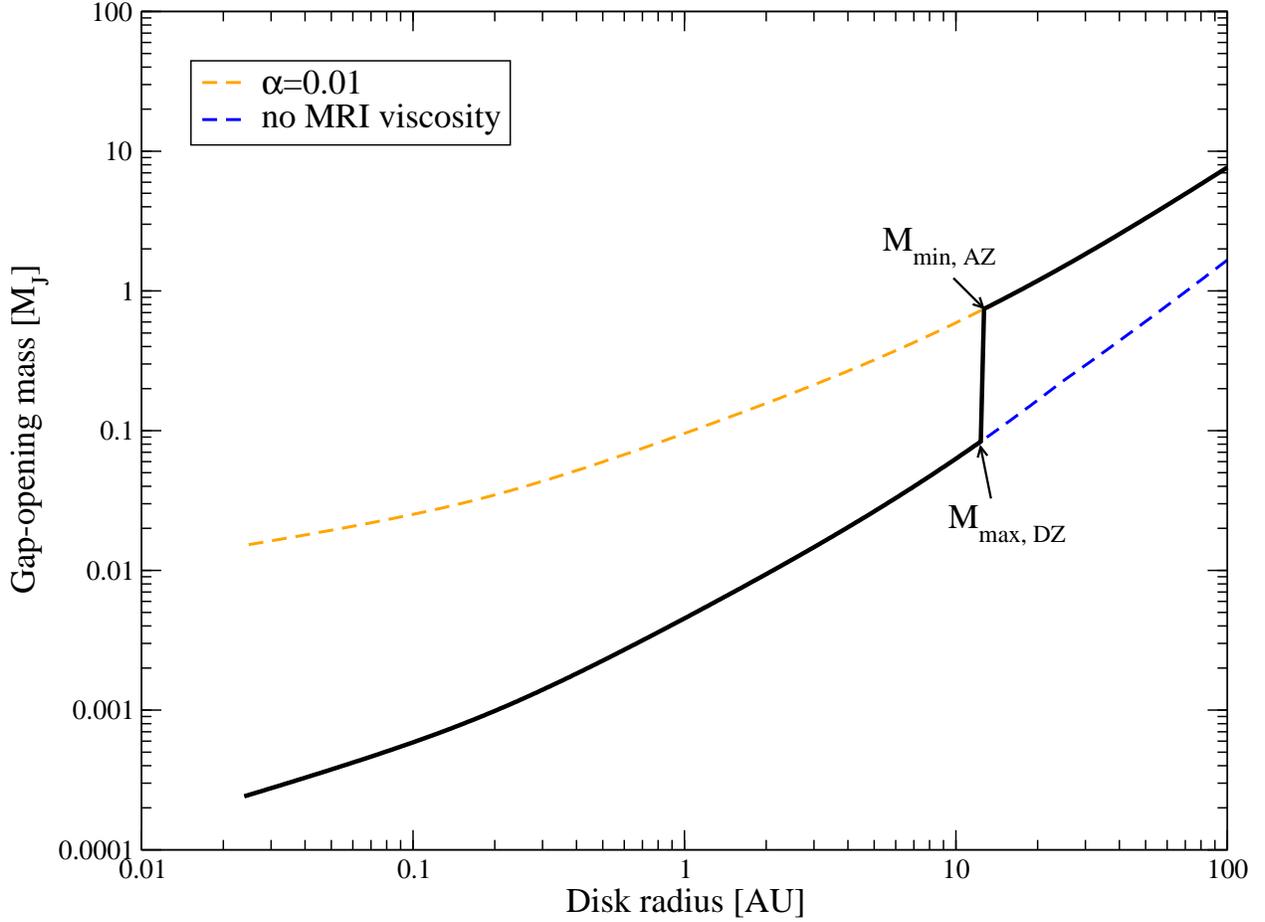} \caption[Gap-opening mass in a disk
inside a clustered environment]{Gap-opening masses for a disk
exposed to an external star which has the surface mass density, at 1
AU, of \(\Sigma_{0}=10^3 \ {\rm g \ cm^{-2}}\). The lower dashed
line shows the gap-opening mass for the case of no MRI viscosity,
while the upper dashed line shows the gap-opening masses for
\(\alpha_{ss}=10^{-2}\).  For the magnetic Reynolds number
\(Re_{M}=10^3\), the fiducial minimum gap-opening mass throughout
the disk is shown in a solid line. \label{fig2}}
\end{figure}

\begin{figure}
\plottwo{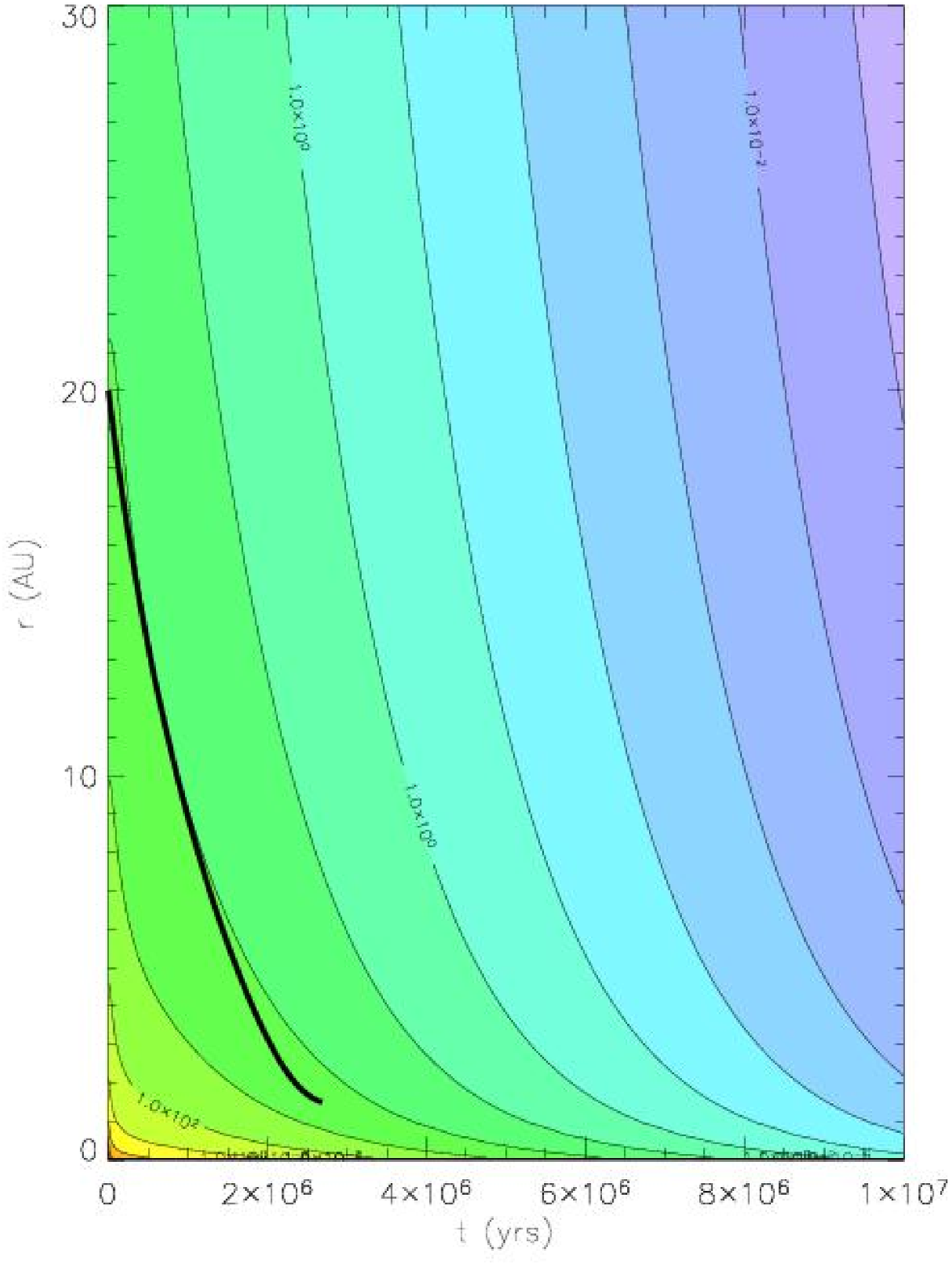}{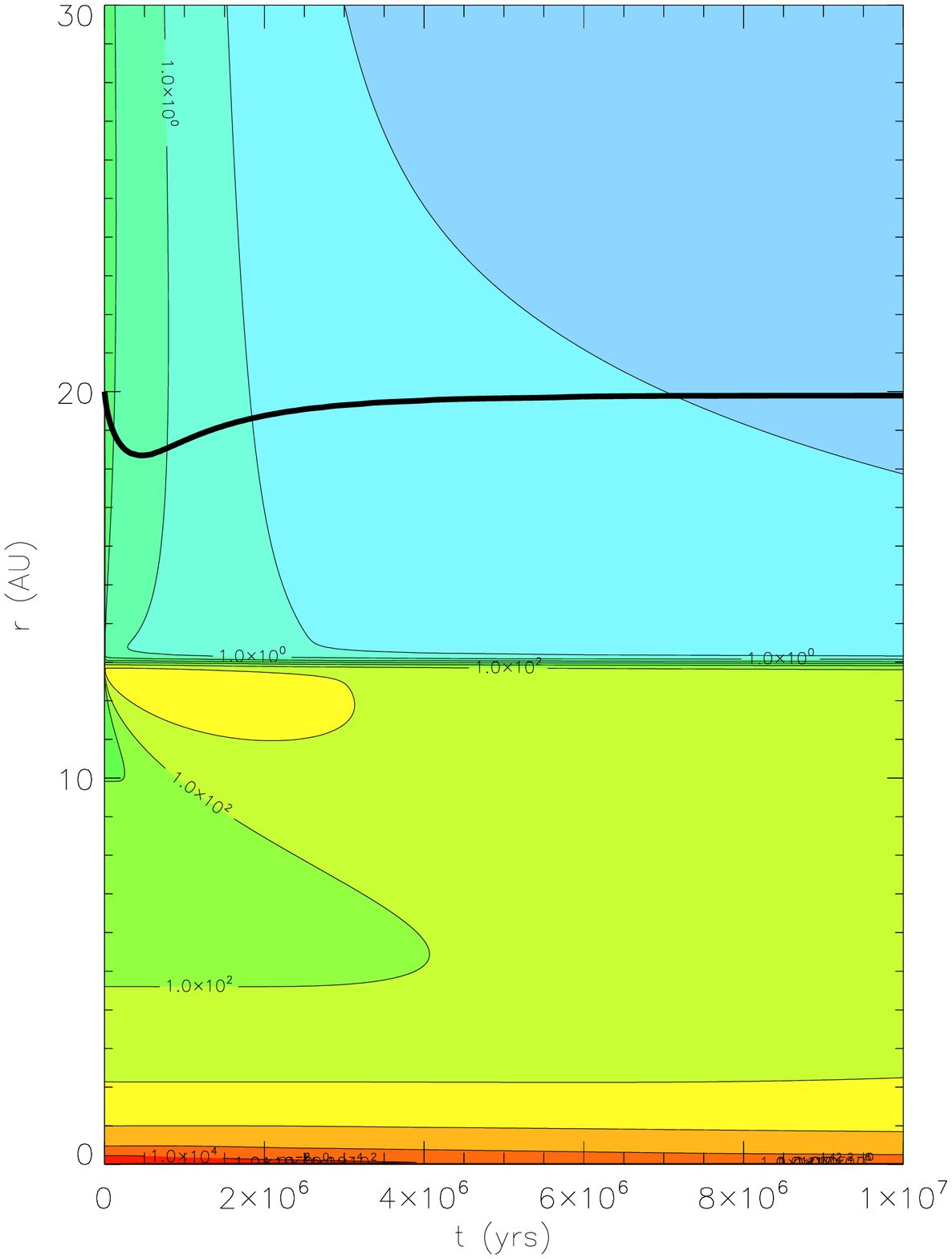}
%\unitlength1cm
%\begin{minipage}[t]{18.5cm}
%\hspace{0.5cm}
%\begin{minipage}[h]{14cm}
%\begin{picture}(5.5,6)
%\includegraphics[width=5cm]{fig4a.eps}
%%\includegraphics[width=5cm]{figs4/test613_1e7.pdf}
%\end{picture}
%\hspace{2cm}
%\begin{picture}(5.5,6)
%\includegraphics[width=5cm]{fig4b.eps}
%%\includegraphics[width=5cm]{figs4/test612_convergence2.eps}
%\end{picture}
%\end{minipage}
%\end{minipage}
%\begin{minipage}[t]{18cm}
%\vspace{0.5cm}
%
\caption[Evolution of a disk with a dead zone: Earth mass planet
inside the dead zone]{Left:Time evolution of the disk and planet
without a dead zone.  The trajectory of a 1 \(M_{E}\) planet,
initially at 20 AU, is shown in a heavy black line.  Also shown is
surface mass density contours which become denser from blue to red.
The planet migrates down to \(\sim 1\) AU within \(3\times 10^6\)
years. Right: Time evolution of the 1D disk with a dead zone's outer
radius at 13 AU. The 1 \(M_{E}\) planet migrates inward a bit, but
then gets repelled by the mass accumulation at the edge of the dead
zone and migrates out to \(\sim 19.8\) AU. Color figures are
available on the web. \label{fig3}}
%\end{minipage}
\end{figure}

\begin{figure}
\plottwo{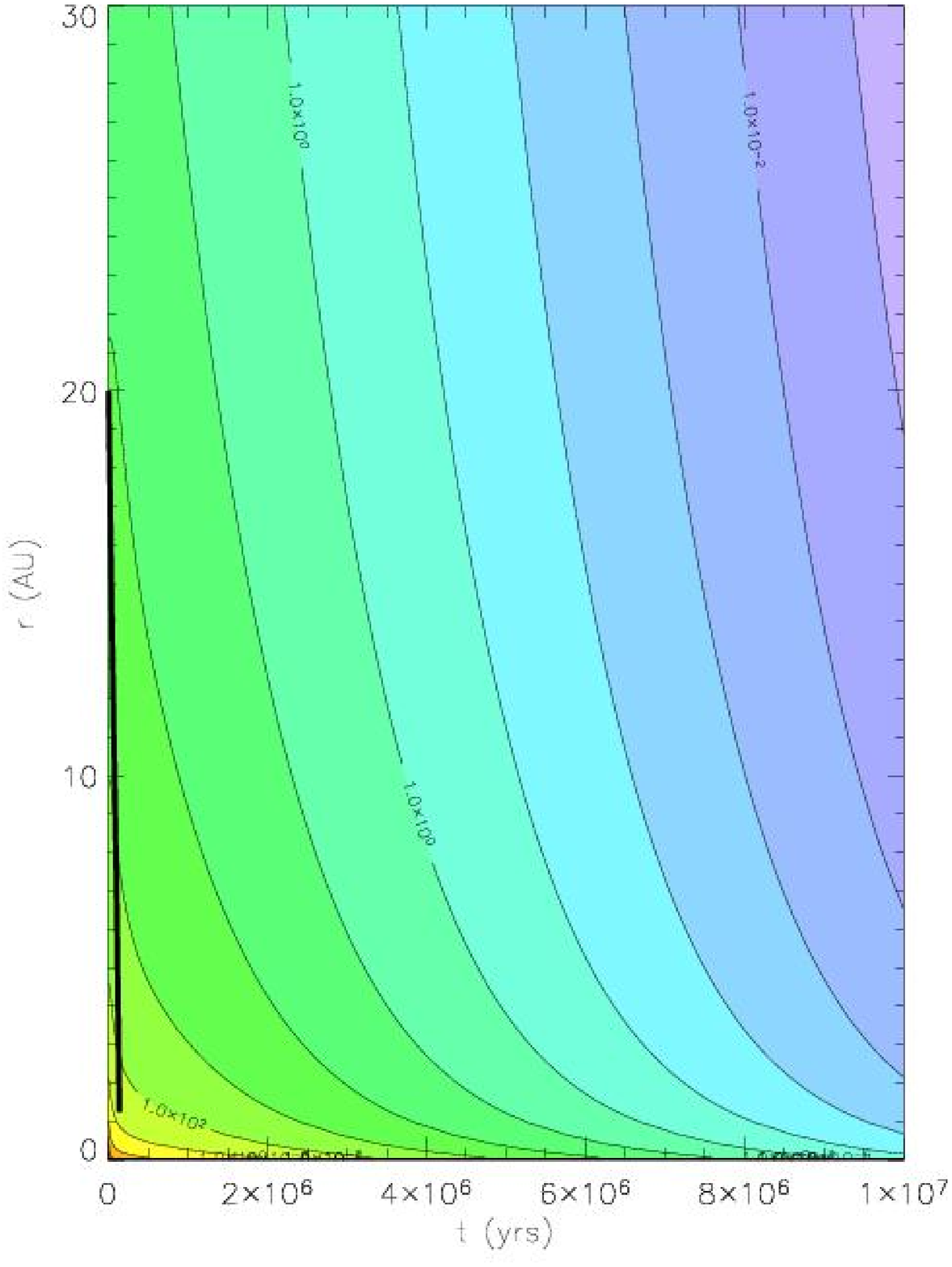}{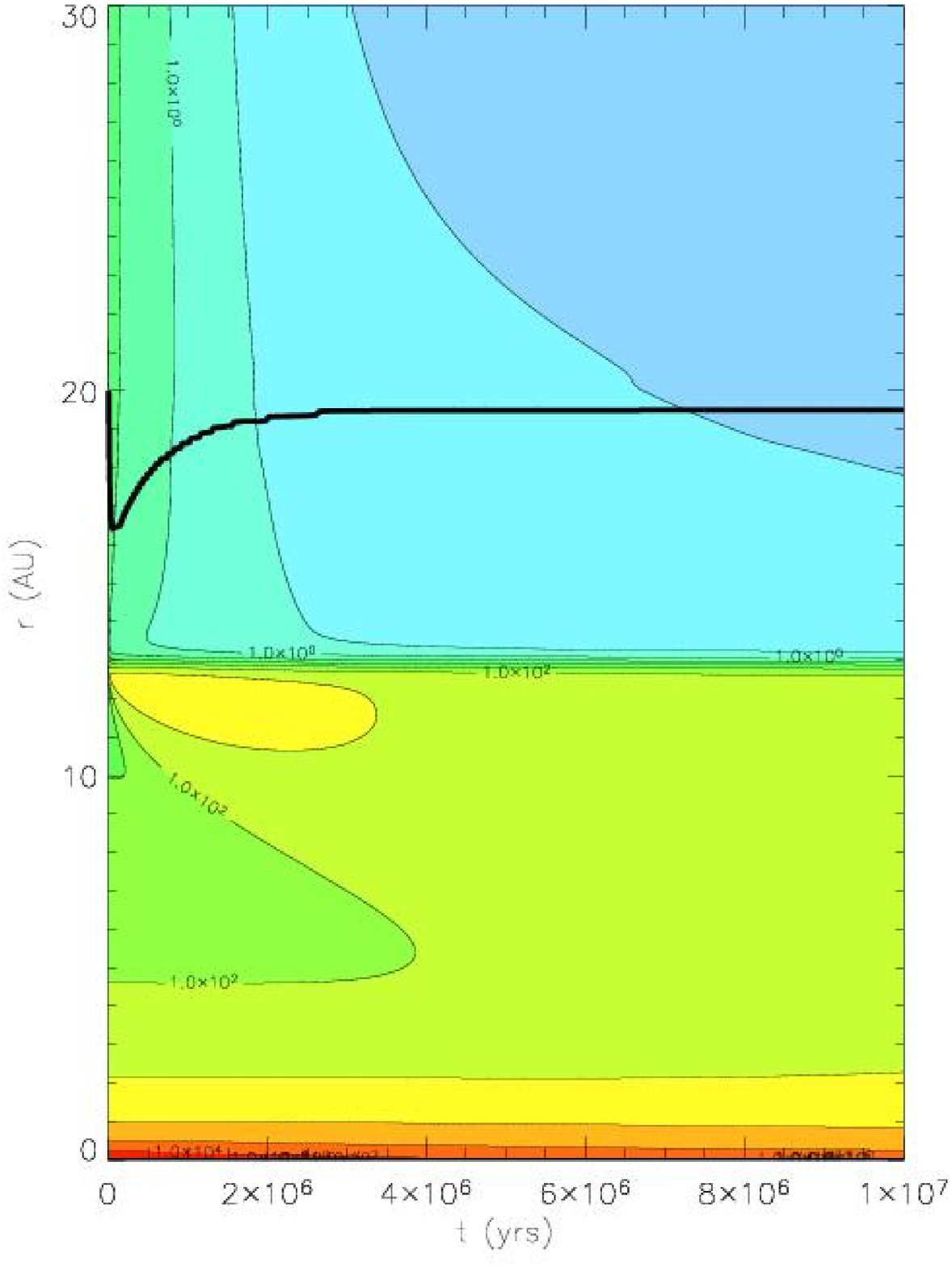}
%\unitlength1cm
%\begin{minipage}[t]{18.5cm}
%\hspace{0.5cm}
%\begin{minipage}[h]{14cm}
%\begin{picture}(5.5,6)
%\includegraphics[width=5cm]{fig6a.eps}
%%\includegraphics[width=5cm]{figs4/test623_1e7.pdf}
%\end{picture}
%\hspace{2cm}
%\begin{picture}(5.5,6)
%\includegraphics[width=5cm]{fig6b.eps}
%%\includegraphics[width=5cm]{figs4/test622_convergence.pdf}
%\end{picture}
%\end{minipage}
%\end{minipage}
%\begin{minipage}[t]{18cm}
%\vspace{0.5cm}
%
\caption[Evolution of a disk with a dead zone: 10 Earth mass planet
outside the dead zone]{Left:Time evolution of the disk and planet
without a dead zone.  The trajectory of a 10 \(M_{E}\) planet,
initially at 20 AU, is shown in a heavy black line.  Also shown is
surface mass density contours which become denser from blue to red.
The planet migrates down to \(\sim 1\) AU within \(1.5\times 10^5\)
years. Right: Time evolution of the 1D disk with a dead zone's outer
radius at 13 AU. The 10 \(M_{E}\) planet migrates inward a bit, but
then gets repelled by the mass accumulation at the edge of the dead
zone and migrates out to \(\sim 19.5\) AU. \label{fig4}}
%\end{minipage}
\end{figure}

\begin{figure}
\plottwo{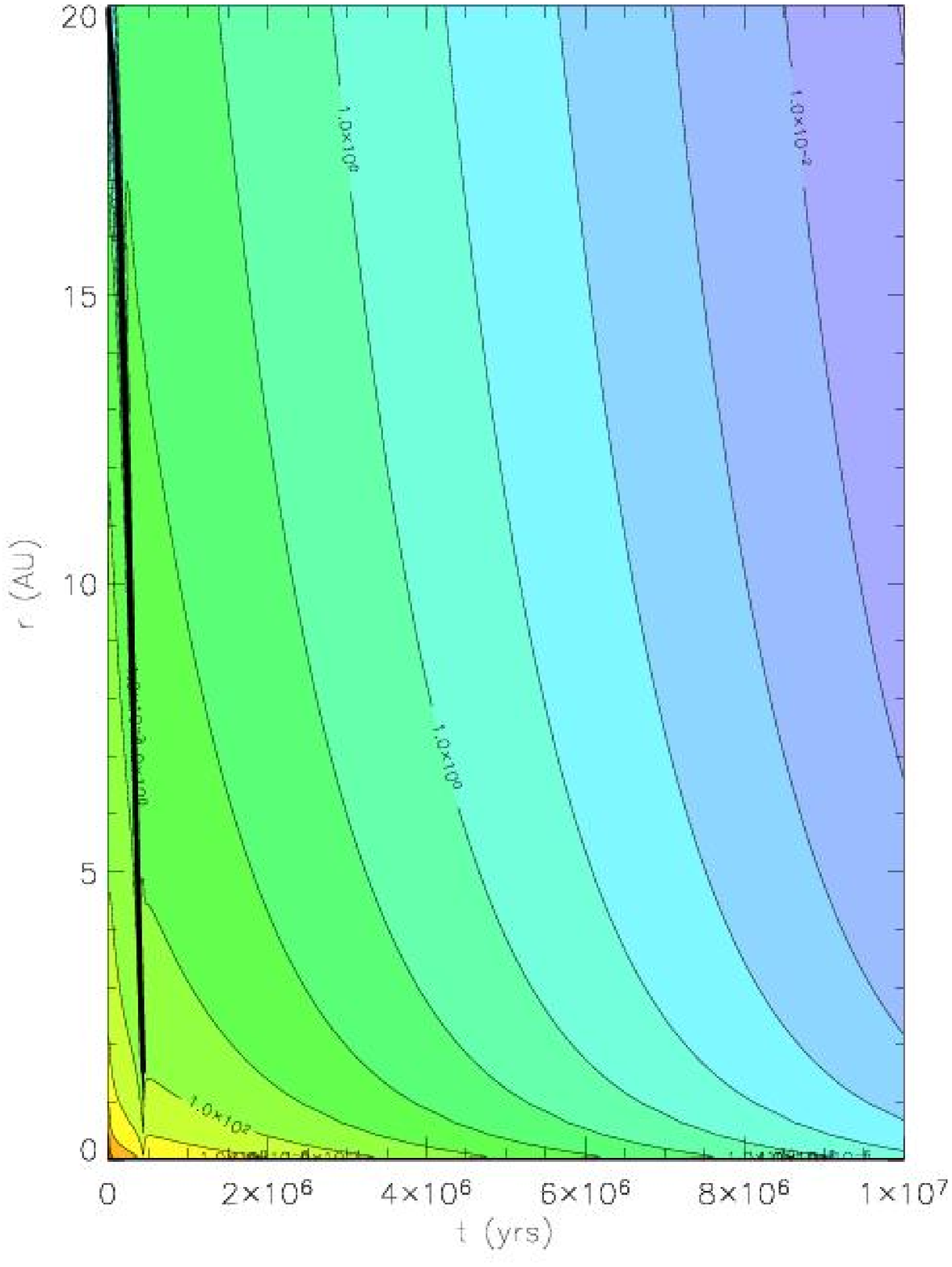}{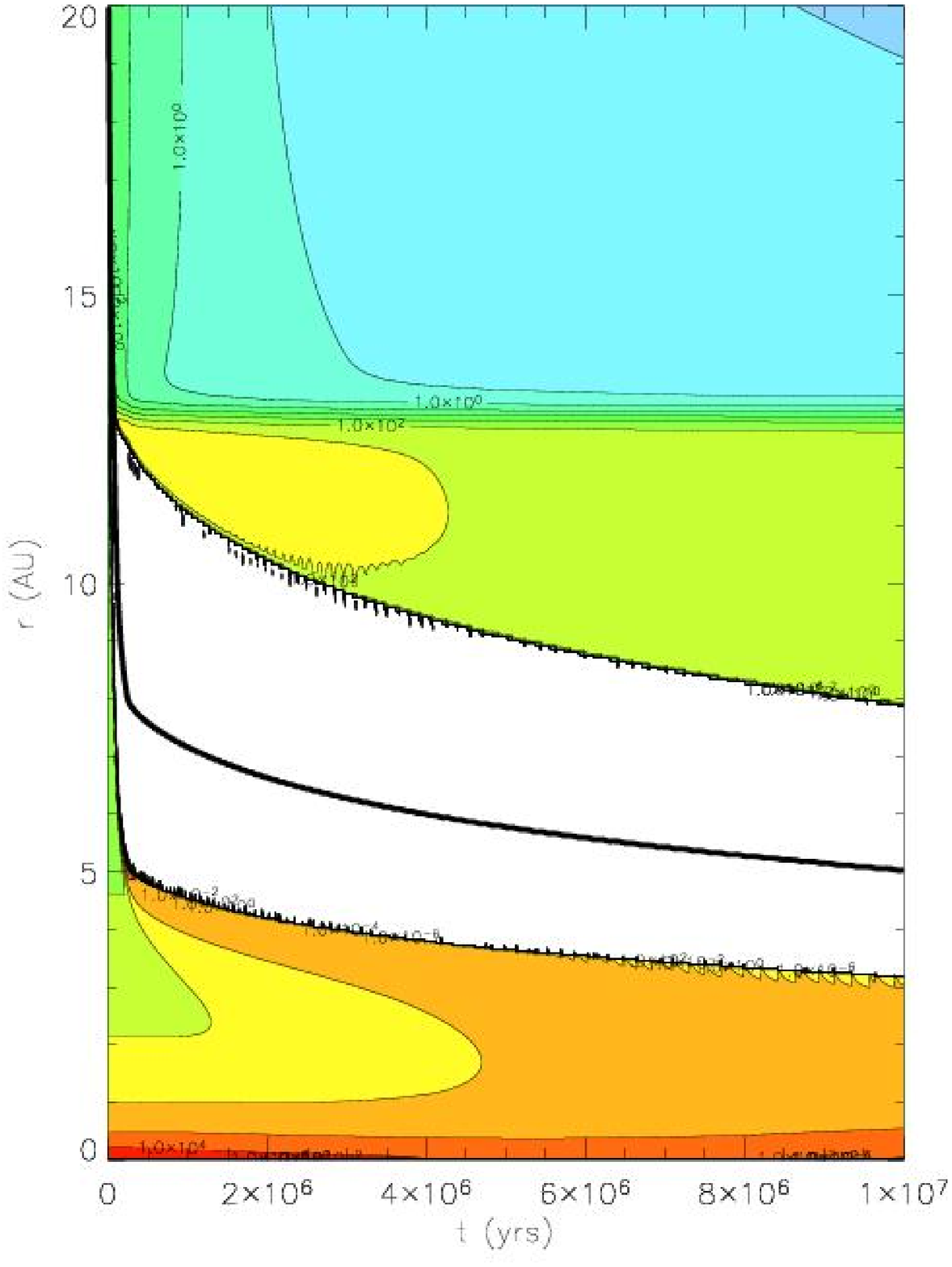}
%\unitlength1cm
%\begin{minipage}[t]{18.5cm}
%\hspace{0.5cm}
%\begin{minipage}[h]{14cm}
%\begin{picture}(5.5,6)
%\includegraphics[width=5cm]{fig7a.eps}
%%\includegraphics[width=5cm]{figs4/test633_1e7.pdf}
%\end{picture}
%\hspace{2cm}
%\begin{picture}(5.5,6)
%\includegraphics[width=5cm]{fig7b.eps}
%%\includegraphics[width=5cm]{figs4/test632_1e7.pdf}
%\end{picture}
%\end{minipage}
%\end{minipage}
%\begin{minipage}[t]{18cm}
%\vspace{0.5cm}
%
\caption[Evolution of a disk with a dead zone: Jupiter mass planet
outside the dead zone]{Left:Time evolution of the disk and planet
without a dead zone. The trajectory of a 1 \(M_{J}\) planet,
initially at 20 AU, is shown in a heavy black line.  Also shown is
surface mass density contours which become denser from blue to red.
The planet migrates inward as it opens a gap and plunges into the
central star within half a million years.  Right: Time evolution of
the 1D disk with a dead zone's outer radius at 13 AU.  The 1
\(M_{J}\) planet migrates inward as it opens a wide gap at around
\(6\) AU.  Its migration is significantly slowed down after the
gap-opening. \label{fig5}}
%\end{minipage}
\end{figure}

\begin{figure}
\plotone{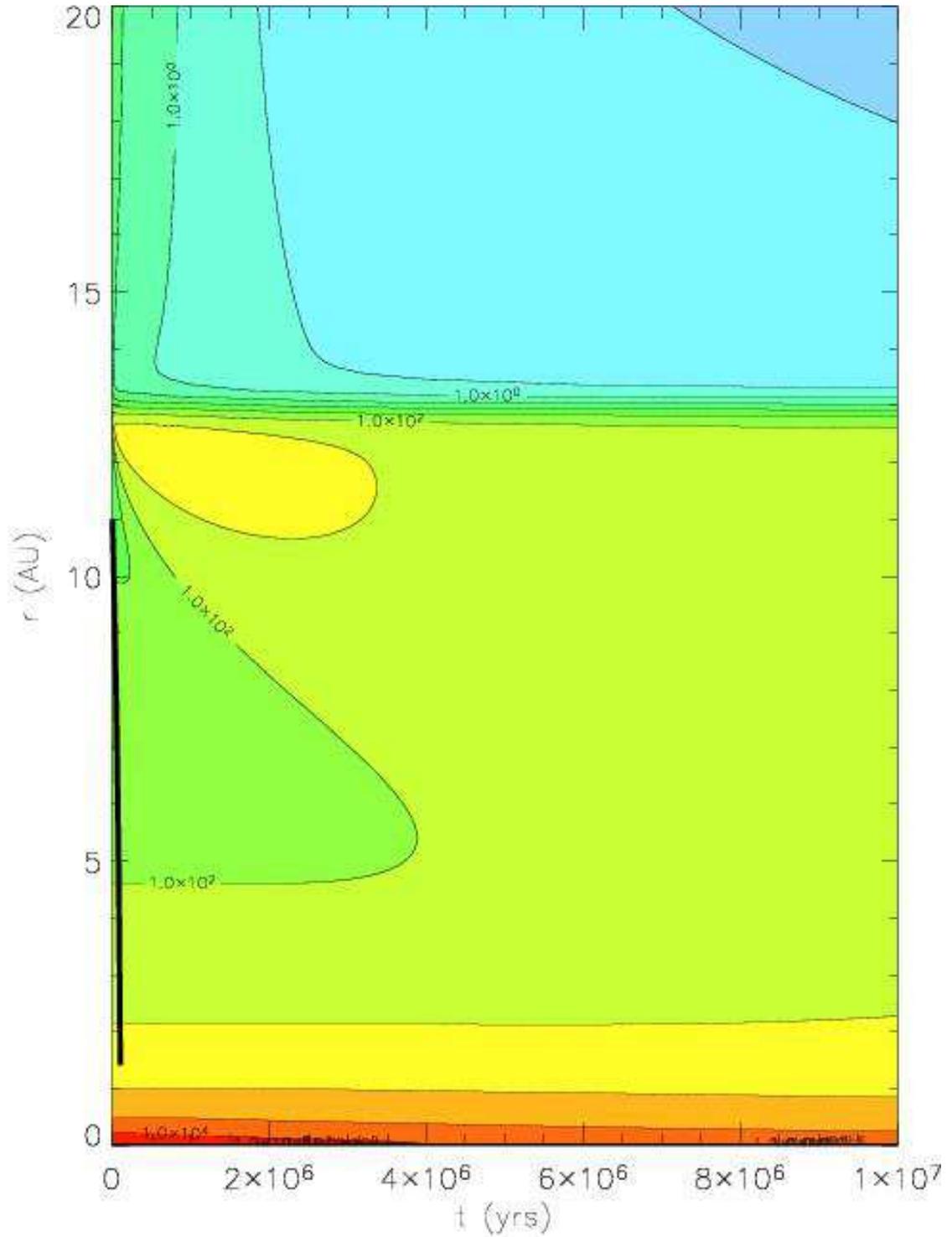}
\caption[Evolution of a disk with a dead zone: Earth mass planet
inside the dead zone]{The same as Fig. \ref{fig3}, but the 1
\(M_{E}\) planet starts migrating from just inside the dead zone,
namely \(11\) AU. In this case, the type I migration becomes even
faster inside the dead zone due to the increase in mass density.
From Fig. \ref{fig2}, the planet is expected to open a gap around
0.7 AU, which is not shown here. \label{fig6}}
\end{figure}

\begin{figure}
\plotone{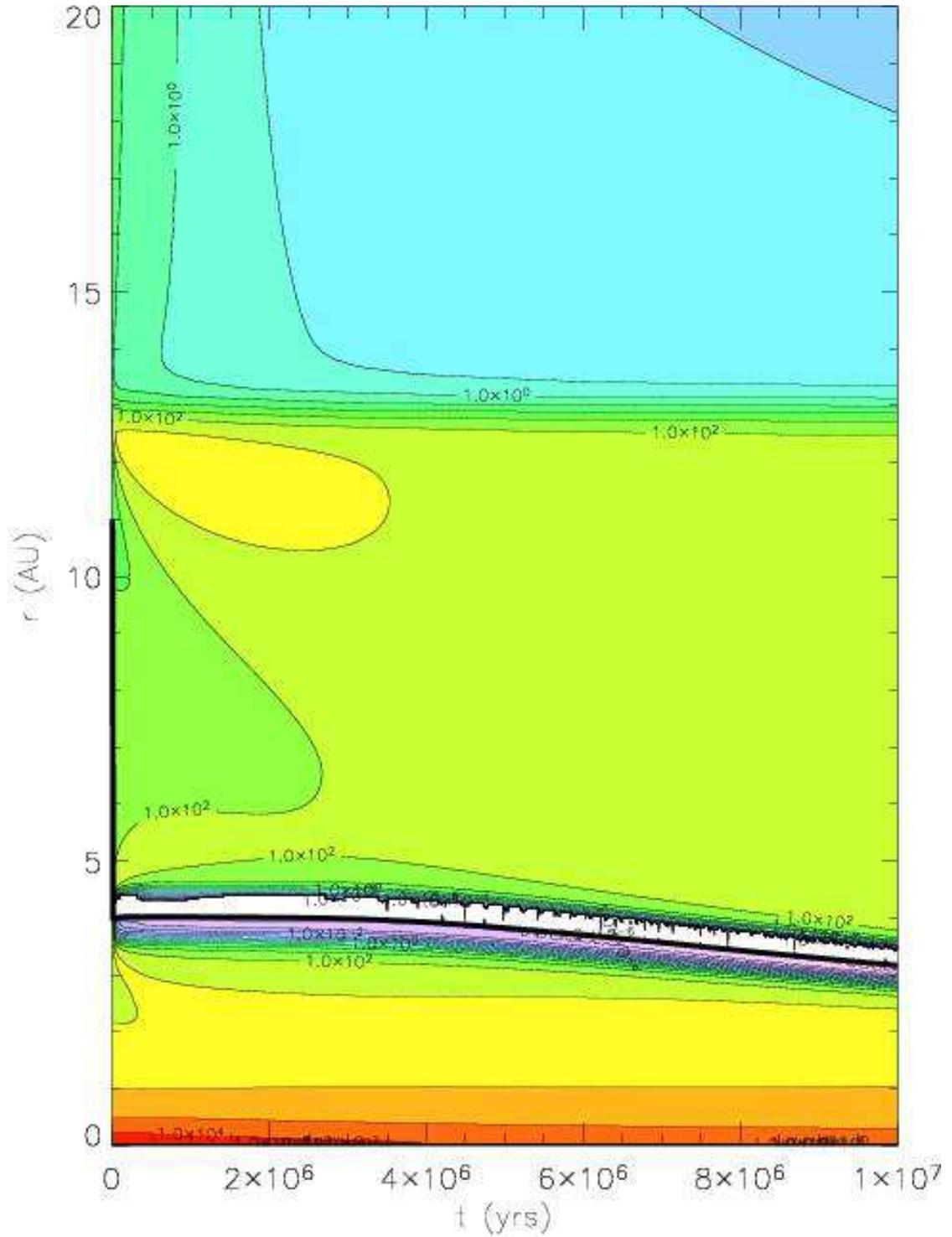}
\caption[Evolution of a disk with a dead zone: 10 Earth mass planet
inside the dead zone]{The same as Fig. \ref{fig4}, but the 10
\(M_{E}\) planet starts migrating from just inside the dead zone,
namely \(11\) AU. The planet migrates inward, and opens a gap at
\(\sim 4\) AU, which agrees well with the estimate from Fig. \ref{fig2} (\(\sim 5\) AU).
After the gap-opening, its migration speed is significantly slowed
down. \label{fig7}}
\end{figure}

\begin{figure}
\plotone{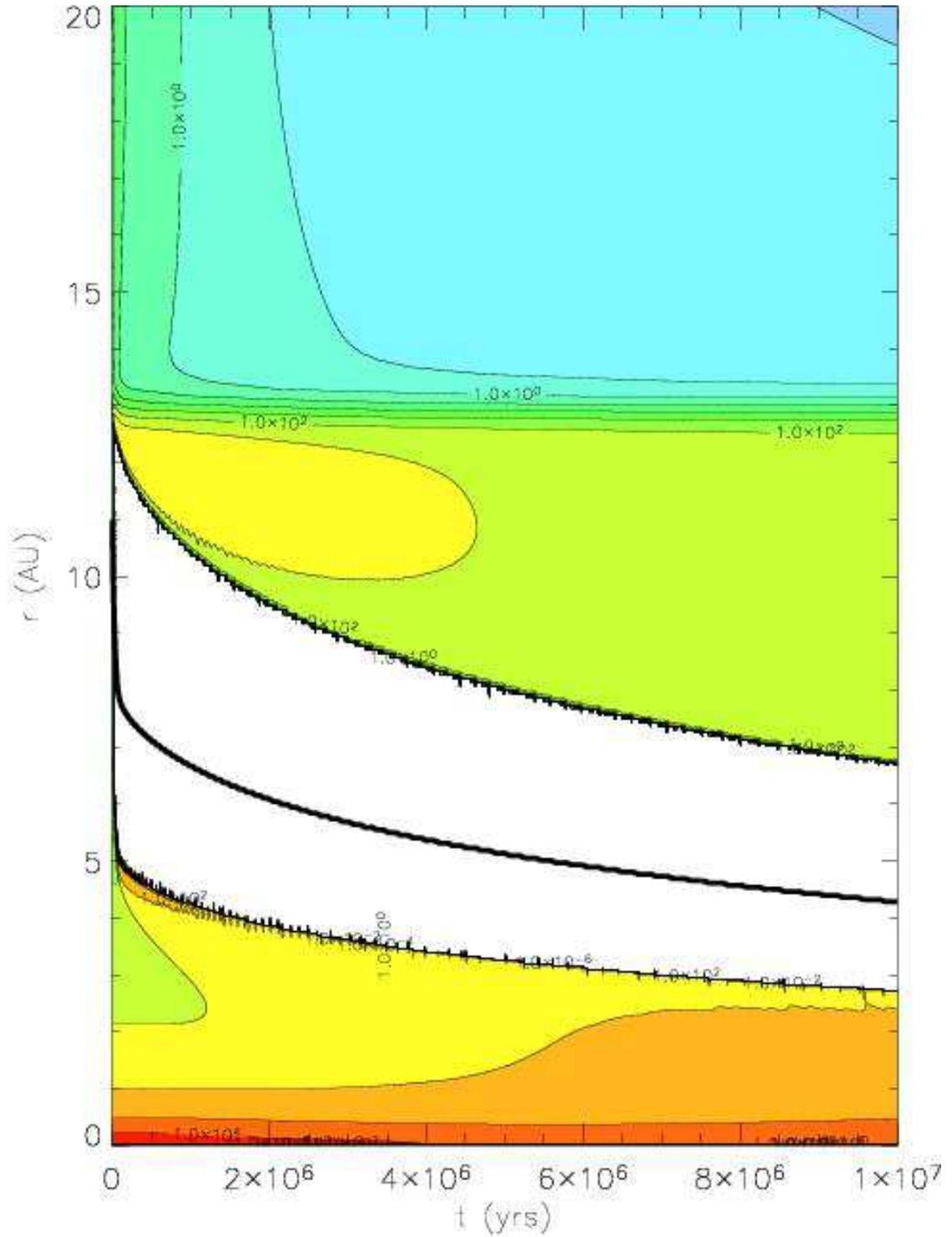}
\caption[Evolution of a disk with a dead zone: Jupiter mass planet
inside the dead zone]{The same as Fig. \ref{fig5}, but the 1
\(M_{J}\) planet starts migrating from just inside the dead zone,
namely \(11\) AU. The planet migrates inward, and opens a gap at
\(\sim 5\) AU.  After the gap-opening, its migration speed is
significantly slowed down. \label{fig8}}
\end{figure}

\begin{figure}
\plotone{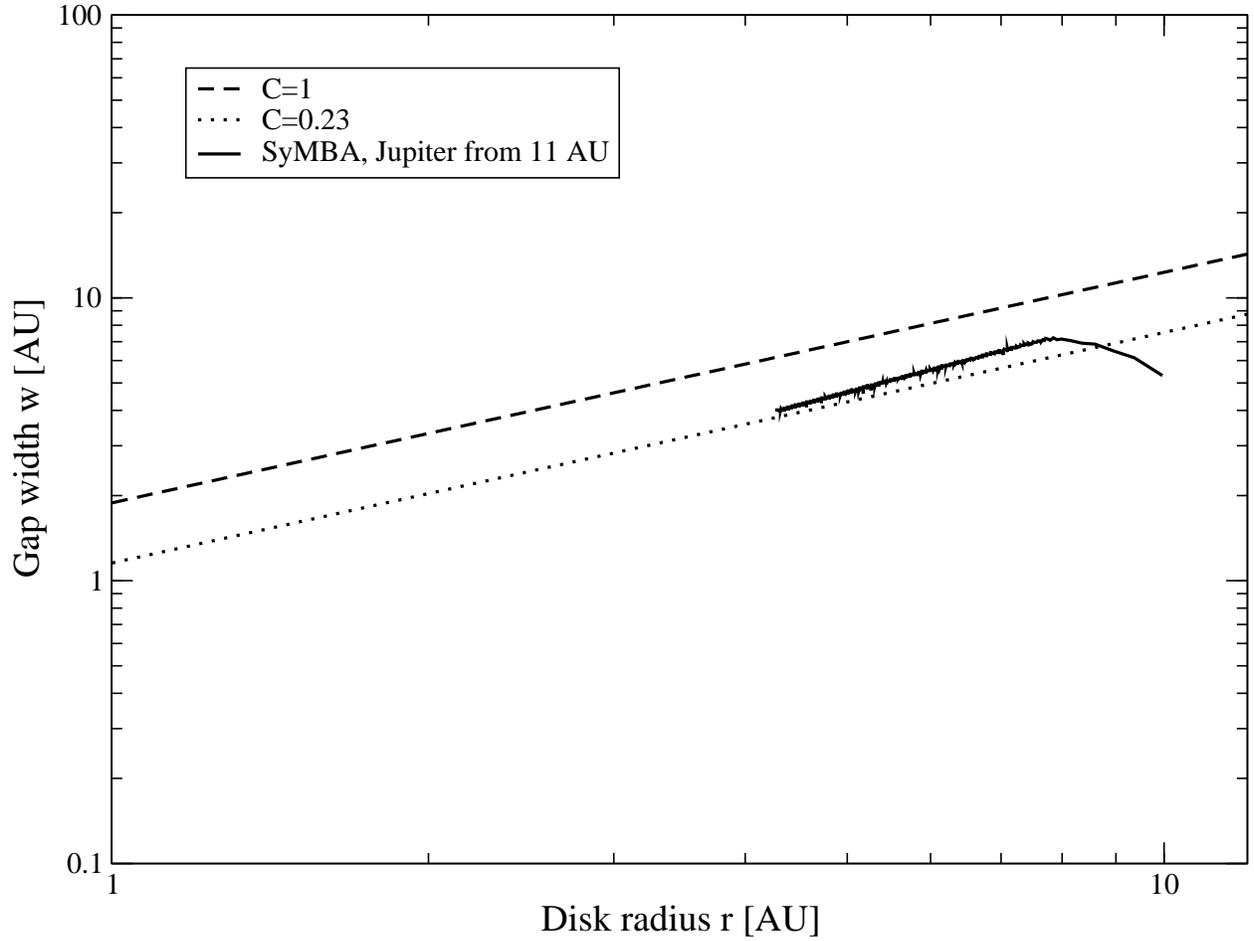}
\caption[Gap width]{The gap width calculated by the Eq. \ref{gapeq}
compared with the one determined from the simulations in Fig.
\ref{fig5}. The dashed and dotted lines correspond to \(C=1\) and
\(C=0.23\) respectively. \label{fig9}}
\end{figure}

\begin{figure}
\plotone{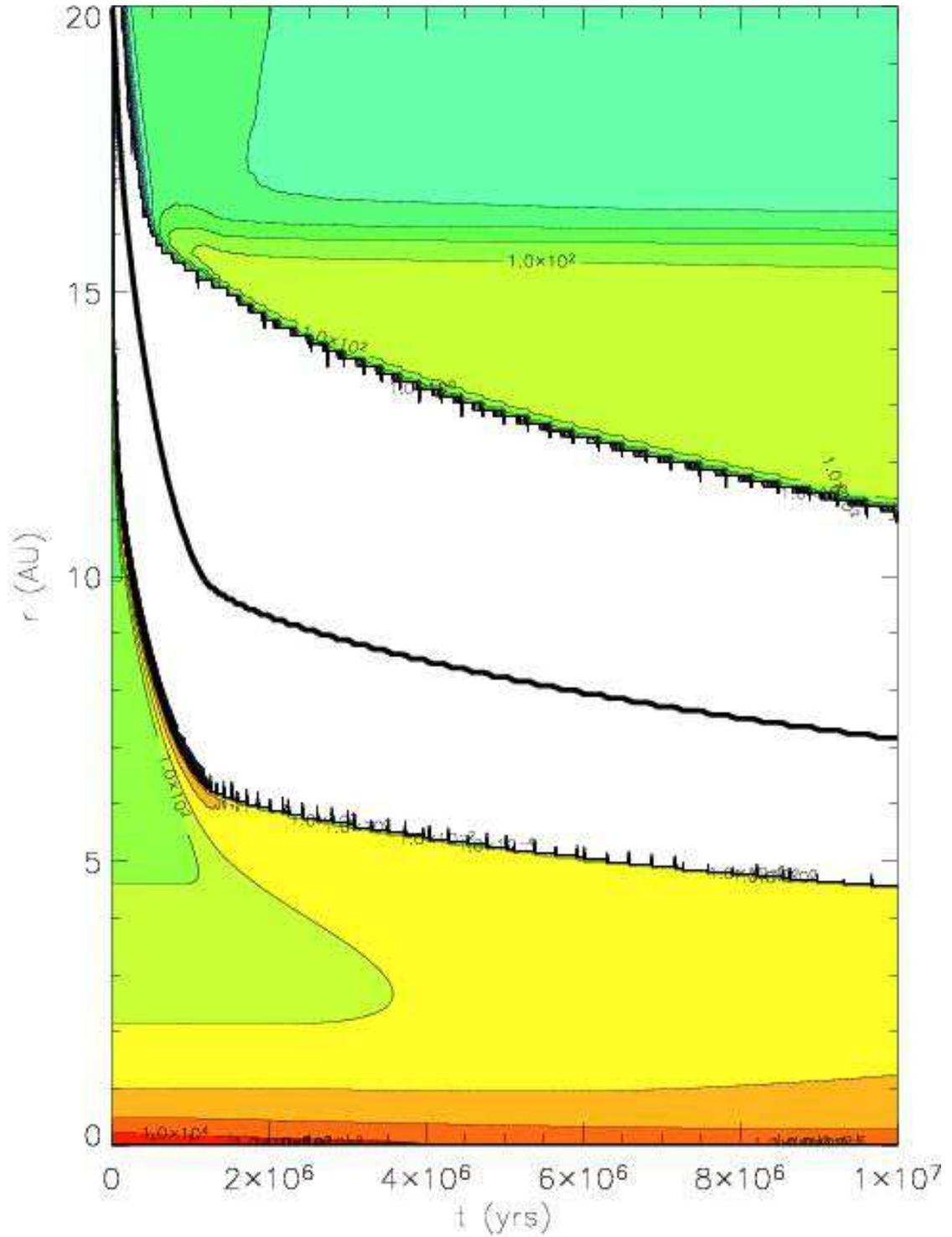}
\caption[Evolution of a disk with a dead zone] 
%(\(\alpha=10^{-5}\) inside and \(\alpha^{-3}\) outside it)]
{The same as Fig.
\ref{fig5}, but the viscosity parameter is \(\alpha=10^{-3}\)
outside the dead zone and \(\alpha=10^{-5}\) inside the dead zone. A
Jupiter mass planet migrates into the dead zone, and slows down
significantly after it opens a wide gap in the dead zone.  The final orbital 
radius of the planet is \(\sim 7\) AU.
\label{fig10}}
\end{figure}

\begin{figure}
\plotone{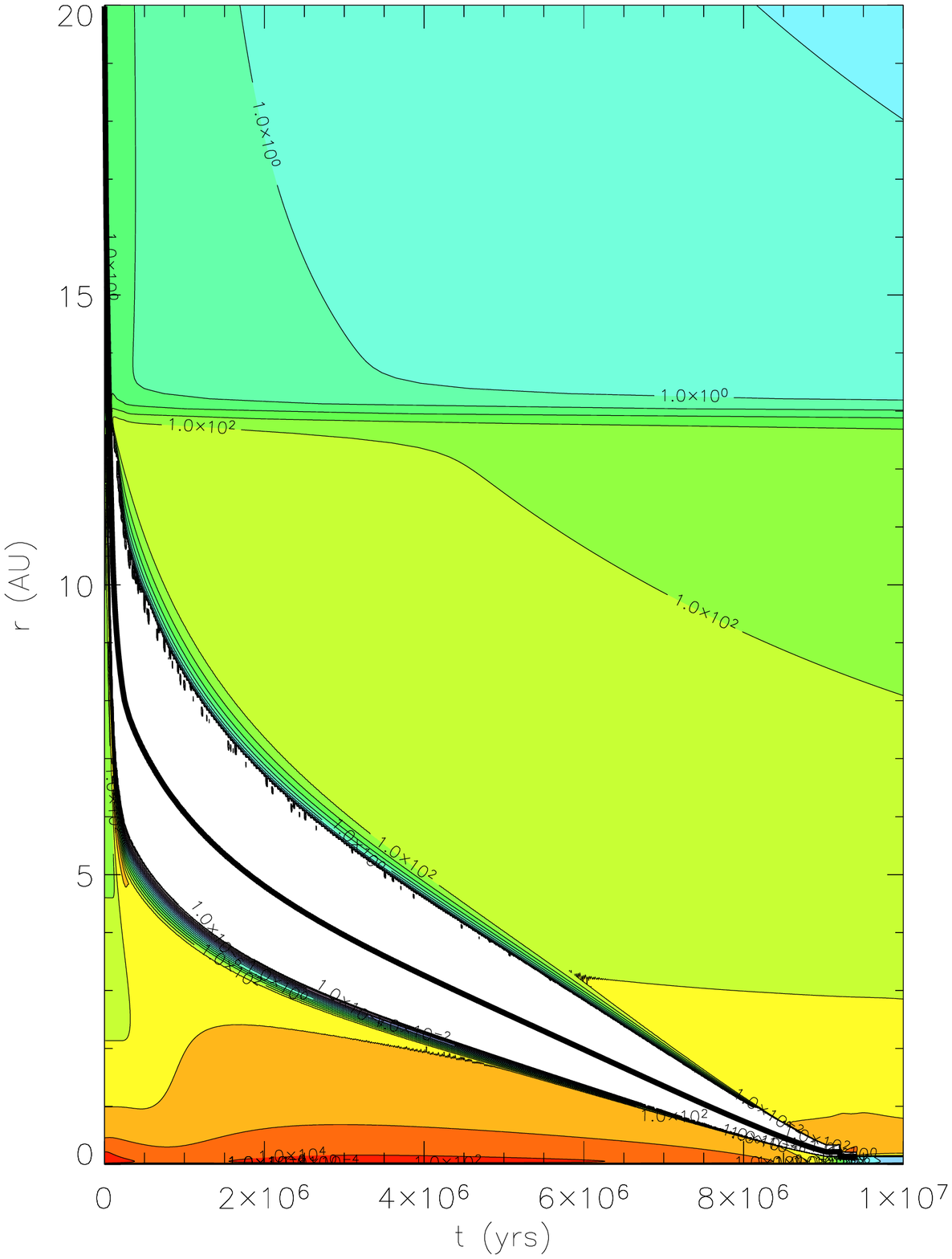}
\caption[Evolution of a disk with a dead zone] 
%(\(\alpha=10^{-4}\) inside and \(\alpha^{-2}\) outside it)]
{The same as Fig.
\ref{fig5}, but the viscosity parameter is \(\alpha=10^{-2}\)
outside the dead zone and \(\alpha=10^{-4}\) inside the dead zone. A
Jupiter mass planet migrates into the dead zone as it opens a gap.  
The final orbital radius of the planet is \(\sim 0.1\) AU. \label{fig11}}
\end{figure}

\begin{figure}
\plotone{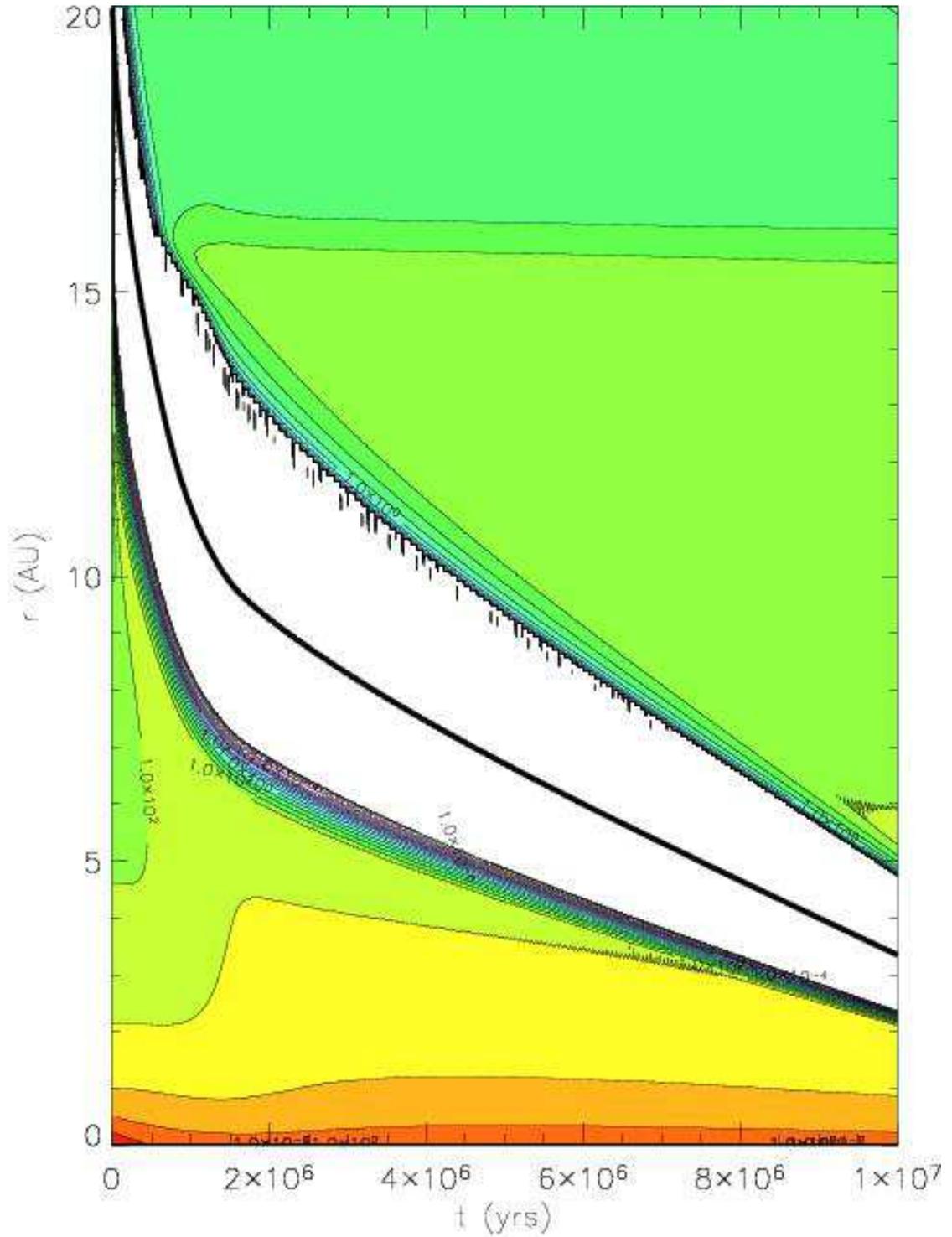}
%\includegraphics[width=5cm]{fig15.eps}
%\includegraphics[width=5cm]{figs4/test663_1e7.eps}
%\caption[Evolution of a disk with a dead zone (\(\alpha=10^{-4}\)
%inside and \(\alpha^{-3}\) outside it)]
\caption[Evolution of a disk with a dead zone]
{The same as Fig.
\ref{fig5}, but the viscosity parameter is \(\alpha=10^{-3}\)
outside the dead zone and \(\alpha=10^{-4}\) inside the dead zone. A
Jupiter mass planet migrates into the dead zone as it opens a gap,
but is not slowed as much as Fig. \ref{fig5}. The final orbital radius of the 
planet is \(\sim 3.5\) AU. \label{fig12}}
\end{figure}

%\subsection{Type I migration toward a density jump}
\label{density jump}
%
%As we see in the last section, disks with a dead zone quickly
%develop a density jump at the edge of the dead zone.
Here, we wish to investigate the effect of a jump in
azimuthally-averaged gas disk surface density on the gravitational
torques experienced by embedded non-gap opening (i.e. Type I)
bodies. We introduce an artificial jump into a power-law surface
density profile as follows:
 \begin{equation}
 \Sigma = \Sigma_{\rm base}+ (F-1) \Sigma_{\rm base} \left
[\frac{1}{2}-\frac{1}{2} \tanh \left ( \frac{r-r_{\rm edge}}{\omega}
\right ) \right ]
 \label{Sigma with step} \ ,
\end{equation}
which is a good approximation of the density jump seen in our
simulations, for example, Fig. \ref{fig1}.  Thus we have a surface
density which, going from large to small $r$, rises smoothly by a
factor F over a radial distance $\omega$, centered on $r=r_{\rm
edge}$.
%We experiment with interpolating
%The radial pressure gradient in the disk causes both the epicyclic and
%the azimuthal frequency to differ from their Keplerian values:
The radial pressure gradient in the disk causes the azimuthal
frequency to differ from their Keplerian values:
\begin{equation}
\Omega^2(r) = \frac{G M_*}{r^3} + \frac{1}{r \rho}
\frac{\partial}{\partial r}(\rho c_s^2) \label{pressure supported
omega} \ .
\end{equation}
This then also has a direct effect on the epicyclic frequency since
\begin{equation}
\kappa^2(r) = \frac{1}{r^3} \frac{\partial}{\partial r}\left ( r^4
\Omega^2 \right ) \label{pressure supported kappa}.
\end{equation}
Thus, at locations of steep density gradients, these frequencies can
be significantly non-Keplerian.

A planet on a circular orbit in a protostellar disk acts as a
perturbation potential of the disk and exerts torques at the
vicinity of the Lindblad resonances, which lead to the transfer of
angular momentum. The perturbation potential can be expanded into a
Fourier series, and the torque around an \(m\)th order resonance,
\(T_{{\rm tidal},m}\), can be expressed in terms of the amplitude of
the potential of a corresponding order \citep{Goldreich80}. The
planet-disk interaction is calculated through the average torque per
radial step (i.e. torque density \(dT_{\rm tidal}/dr\)) as can be
seen in Eq. \ref{surfevol}.
%Instead of summing up all angular momentum flux from all orders of Lindblad resonances
Assuming the angular momentum transfer is smooth (i.e. the damping
distance of the waves is longer than the distance between
resonances), a smooth torque density can be written as
\citep{Ward97}:
\begin{equation}
\frac{dT_{\rm tidal}}{dr}=T_{{\rm tidal},m} \left| \frac{dm}{dr}
\right| \ .
\end{equation}
%
%as in \cite{Ward97}.

Following the formulation of \cite{Ward97} \citep[see
also][]{Menou04}, the smooth torque density experienced by the disk
due to a planet of mass $M_p=\mu M_*$ on a circular orbit about a
primary of mass $M_*$ at radius $r_p$, is
\begin{equation}
\frac{dT_{\rm tidal}}{dr}(r) = {\rm sgn}(r-r_p) \frac{2 \mu^2
\Sigma(r) r_p^4 \Omega_p^4}{r(1+4 \xi^2)\kappa^2} m^4 \psi^2
\label{torque density}
\end{equation}
where $\xi \equiv m c_s/r \kappa$ measures the change to the Lindblad resonances due to finite gas pressure, $c_s$ is the gas sound speed, and $\psi$ is the dimensionless forcing function due to a planet's
potential that is written as:
\begin{equation}
\psi = \frac{\pi}{2} \left [ \frac{1}{m} \left | \frac{d
b^m_{1/2}}{d \beta} \right | + 2 \frac{\Omega}{\kappa} \sqrt{1 +
\xi^2} \ b^m_{1/2}(\beta)\right] \label{forcing func},
\end{equation}
with $\beta \equiv r/r_p$ and $b^m_{1/2}(\beta)$ being the Laplace
coefficient:
\begin{equation}
b^m_{1/2}(\beta) \equiv \frac{2}{\pi} \int^{\pi}_0 \frac{\cos m
\theta d \theta}{\sqrt{1 - 2 \beta \cos \theta + \beta^2}} \ .
\end{equation}
The largest contribution to the integral above comes from \(\theta
\ll 1\), and it can be well approximated by the modified Bessel
function of the second kind of order 0, \(K_0(z)\), which decays
exponentially for \(z\gg 1\) \citep{Goldreich80,Menou04}.

To take account of the 3D torque effect (i.e. effect of the finite
thickness of the disk) in a simple approximate way, we followed
\cite{Menou04} with a softened approximation,
%Following \cite{Menou04},
%%\citet*{2004ApJ...606..520M},
%we account in a simple, approximate way for the finite thickness of the
%disk by replacing the Laplace
%coefficient with a softened approximation,
\begin{equation}
b^m_{1/2}(\beta) \approx \frac{2}{\pi \beta^{1/2}}K_0 \left ( m
\sqrt{\beta - 2 + \frac{1}{\beta} + \frac{(B h)^2}{r r_p}} \right )
\label{softened laplace coeff}
\end{equation}
with $h \approx c_s/\Omega$ the disk scale height, and $B$ a
dimensionless scaling factor parameterizing the softening in units
of the disk scale height. Therefore, although the hybrid code
calculates a 1D disk evolution, the interaction between the disks
and planets is evaluated by using an approximated 3D torque density.
In going from a summation of torques at discrete resonances to a
torque density, the wavenumber $m$ is turned into a continuous
function of radius:
\begin{equation}
m(r) = \left [ \frac{\kappa^2}{(\Omega-\Omega_p)^2-c_s^2/r^2}\right
]^{1/2}.
\end{equation}
%
%Eq. \ref{torque density} implies that the reaction torque density
%from the inner disk to the planet is always positive since
%\(r<r_p\), while the reaction torque density from the outer disk is
%always negative since \(r>r_p\). The type I planet migration speed
%is obtained by summing over the reaction torque density:
%\begin{equation}
%v_{\rm typeI}=\frac{2}{M_pr_p\Omega_p}\int-\left(\frac{dT_{\rm
%tidal}}{dr}\right) \ dr \ .
%\end{equation}
%%
%This value is usually negative (i.e. planets migrate inward) because
%(i) epicycle frequency \(\kappa\) in Eq. \ref{torque density} is a
%decreasing function of disk radius, which is usually steeper than
%the surface mass density \(\Sigma\), and (ii) outer resonances feel
%a stronger forcing function \(\psi\), and therefore produce a larger
%torque in magnitude.

Using Eq. \ref{Sigma with step} for the surface density, we
numerically compute the summed torque from Eq. \ref{torque density}
in Matlab.    We normalize the torque as in \cite{Ward97}, dividing
by a reference torque
\begin{equation}
\tau_0=\mu^2 \left ( \frac{\pi r_p^2 \Sigma_p }{M_*} \right ) M_*
(r_p \Omega_p)^2 \left ( \frac{r_p}{h_p} \right )^3,
\end{equation}
where the subscript $p$ denotes a quantity evaluated at $r=r_p$.

\begin{figure}
\plotone{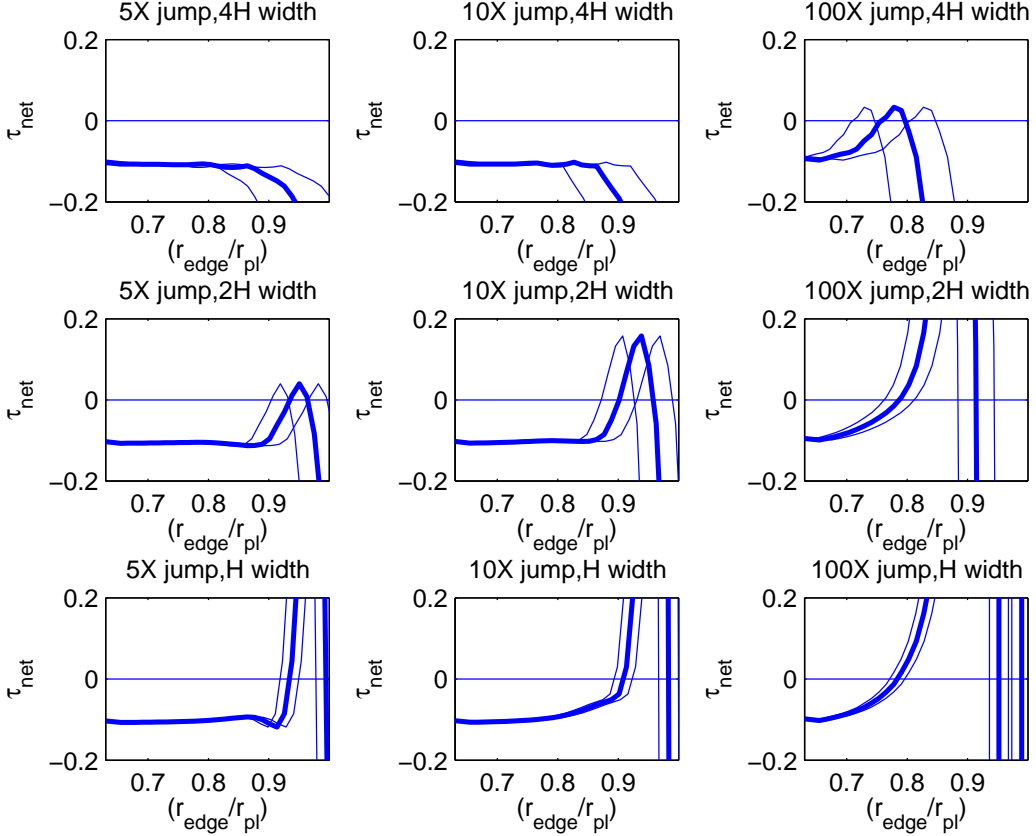}
% source
\caption[Normalized net torque on a planet in a disk with a density
jump (\(B=0\))]{Normalized net torque $\tau_{\rm net}$ density on a
planet embedded in a power-law gas disk containing a jump in
surface.  The planet has orbital radius $r_{\rm pl}$, and the net
torque felt by the planet is plotted as a function of the midpoint
of the density jump, $r_{\rm edge}$.  Away from the edge, the gas
disk surface density is a power law, $\Sigma \propto r^{-3/2}$.  The
calculation is carried out for density jump factors $F=5,10,100$,
occurring over a radial width $\omega=h,2h,4h$ where $h$ is the disk
scale height at $r_{\rm edge}$ (see Eq. \ref{Sigma with step}).  The
torques are computed two-dimensionally, i.e. $B=0$ in Eq.
\ref{softened laplace coeff}. The thick curve shows the net torque,
while the horizontal distance between the thin curves on either side
shows the corresponding radial width $\omega$ of the density jump.
Since $\omega$ is scaled to $h$ at $r_{\rm edge}$, this changes as
the edge midpoint changes. } \label{fig13}
\end{figure}

We begin by adopting $B=0$, i.e. the torques are calculated
two-dimensionally. Fig. \ref{fig13} shows the normalized total torque
computed as a function of $r_{\rm edge}/r_{p}$, for different values
of density jump height $F$ and width $\omega$. The width is in units
of the disk scale height $h$ at $r=r_{\rm edge}$. The base surface
density is a power law, $\Sigma_{\rm base} \propto r^{-3/2}$, as is
the sound speed, $c_s \propto r^{-1/4}$.
%\begin{figure}
%\plotone{figs/h=0_ngrid=1000.eps}
%% source
%\caption{Normalized net torque $\tau_{\rm net}$ density on a planet
%embedded in a power-law gas disk containing a jump in surface.  The
%planet has orbital radius $r_{\rm pl}$, and the net torque felt by the
%planet is plotted as a function of the midpoing of the density jump,
%$r_{\rm edge}$.  Away from the edge, the gas disk surface density is a
%power law, $\Sigma \propto r^{-3/2}$.  The calculation is carried out
%for density jump factors $F=5,10,100$, occurring over a radial width
%$w=h,2h,4h$ where $h$ is the disk scale height at $r_{\rm edge}$ (see
%Eq. \ref{Sigma with step}).  The torques are computed
%two-dimensionally, i.e. $B=0$ in Eq. \ref{softened laplace coeff}}.
%\label{h=0_ngrid=1000}
%\end{figure}
While the planet is still far from the density jump, the net torque
is negative and approximately constant  with $r_{\rm edge}/r_{p}$.
When the surface density jumps by a factor $F=5$ or 10 over a radial
distance of $4 h$, the net torque on the planet is always negative,
no matter where inside its orbit the density jump occurs.  However,
when the density jumps by a factor of 100 over $4h$, the net torque
becomes positive when $r_{\rm edge} \approx 0.75 r_{p}$.  Though
$\tau_{\rm net}$ becomes negative again for slightly larger $r_{\rm
edge}$, a body migrating toward the edge will stop where $\tau_{\rm
net}$ first becomes zero, and get hung up there.  In this way, a
jump in the azimuthally-averaged  gas disk surface density can
%act as a proverbial brick wall for
completely stop Type I migration. 

Note that decrease/fluctuations in 
\(\tau_{\rm net}\) for \(r_{\rm edge}>0.75 r_p\) result because the region
of highest torque density close to the planet is beginning to overlap the
region of the density jump. In reality, this close to the jump, corotation
torques would start to dominate over Lindblad torques, so our calculation of
\(\tau_{\rm net}\) is any case no longer valid here.
Also, it should be pointed out that the steepest jump ($F=100$, $\omega=h$ over a distance $h$) actually
produces a locally imaginary epicyclic frequency and as such would
be subject to Rayleigh instability. This would reduce the density
jump, which may even change the migration direction. We leave this
investigation to future work.
%Such a steep density jump would thus occur at most as a transient state.}

Further examination of the remaining panels of Fig. \ref{fig13}
reveals that when the density jump occurs over $2 h$ or $h$, the
torque eventually becomes positive no matter whether the density
jumps by a factor of 100, 10 or even 5.  The exact value of $r_{\rm
edge}/r_p$ at which this happens changes, but in all cases, it is
greater than the half-width of the jump, $(\omega/2)/r_p$.  In other
words, the net torque becomes zero before the density jump itself
reaches the planet's orbital radius. This is important because once
a planet finds itself at the same radius as the density jump, the
corotation torques responsible for type III migration (see \S 1),
not included in this calculation, may overpower the Lindblad
torques, with the effect being in the opposite sense: the planet is
pulled {\it toward} the higher-density region \citep{Masset06}.
%(Masset, Morbidelli, Crida \& Ferreira, ApJ 2006).
However, when the Lindblad torque switches from negative to positive
at a significant standoff distance from the density jump already,
approaching planets will be stopped while the corotation torques are
still small compared to the Lindblad torques.   Furthermore, for all
cases with density jumps over a distance $h$, as well as over $2 h$
with a 100 $\times$ jump, the magnitude of the largest positive
torque exceeds that of the negative torque.  This ensures that even
if the corotation torque becomes comparable to the Lindblad torque,
the planet will still be stopped in these cases.
%{\color{red} Unfortunately they sort of beat us to the punch; even
%though their mechanism is different, the distinction may be lost on
%some people.  Since they say in their paper that coorbital torques
%overpower Lindblad torques, we have to point out that in out case of
%the outer edge of the dead zone, the migration is probably stopped well
%before the coorbital torques ever come into play.  }}

We repeat the calculation with a vertically-softened torque,
$B=0.5$. As shown by Menou \& Goodman (2004), this ``pseudo-3D"
torque introduces two major differences relative to the 2D
calculation above. First, the torque asymmetry becomes smaller
because the strong contribution from the parts of the disk near the
planet, just outside of the torque cutoff, are weakened.  Second,
the dependence of the net torque on the steepness of the disk
surface density changes qualitatively.   In the 2D case, in a smooth
power-law disk, the net normalized torque is almost independent of
the power law index. However, with softening, the torque asymmetry
becomes weaker as the surface density becomes steeper.
%Since the density jump represents a region of large negative gradient in $\Sigma$,
Both of these effects ought to conspire to make density jumps even
more effective at arresting Type I migration.

\begin{figure}
\plotone{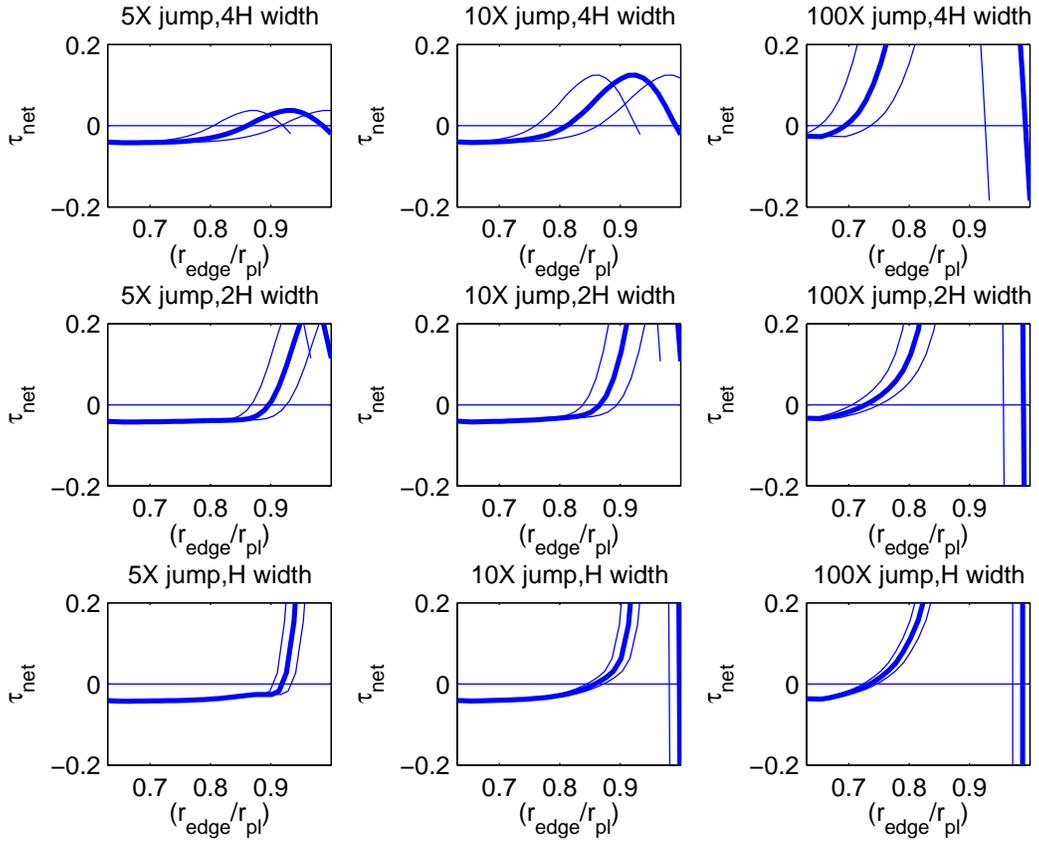}
% source
\caption[Normalized net torque on a planet in a disk with a density
jump (\(B=0.5\))]{The computation shown in Fig. \ref{fig13}, repeated
with $B=0.5$, i.e. vertical softening of torques by half the disk
scale height (Eq. \ref{softened laplace coeff}).} \label{fig14}
\end{figure}

The results of the calculation are shown in Fig. \ref{fig14}. Indeed,
the effect of the surface density jumps on migration is even
stronger than in the 2D case above. Now, every combination of
$F=5,10,100$ and $\omega=h,2h,4h$ results in $\tau_{\rm net}$
becoming positive as $r_{\rm edge}$ approaches $r_{p}$, and in each
case this happens at a somewhat smaller value of $r_{\rm
edge}/r_{p}$ than in the corresponding 2D calculation.  In other
words, the migrating planet is stopped at a slightly greater
standoff distance from the density jump. Also, all combinations
except $F=5$ with $\omega=4 h$ result in  the positive torque far
exceeding the original negative torque well before the planet
reaches the density transition region, thus ensuring that corotation
torques cannot play a significant role.

%In the next section, we will study this effect using time-dependent
%numerical simulations for some typical planetary masses. We used a
%softening parameter \(B=0.2\) in all cases, corresponding to pseudo
%3D disks as already explained.
%\begin{figure}
%\plotone{figs/h=05_ngrid=1000.eps}
%% source
%\caption{The computation shown in Fig. \ref{h=0_ngrid=1000}, repeated
%with $B=0.5$, i.e. vertical softening of torques by half the disk
%scale height (Eq. \ref{softened laplace coeff}).}
%\label{h=05_ngrid=1000}
%\end{figure}
%
%
\bibliography{REF}

\begin{thebibliography}{50}
\expandafter\ifx\csname natexlab\endcsname\relax\def\natexlab#1{#1}\fi

\bibitem[{{Alibert} {et~al.}(2005){Alibert}, {Mordasini}, {Benz}, \&
  {Winisdoerffer}}]{Alibert05}
{Alibert}, Y., {Mordasini}, C., {Benz}, W., \& {Winisdoerffer}, C. 2005, A\&A,
  434, 343

\bibitem[{{Artymowicz}(1993)}]{Artymowicz93b}
{Artymowicz}, P. 1993, ApJ, 419, 166

\bibitem[{{Artymowicz}(2004)}]{Artymowicz04}
{Artymowicz}, P. 2004, in ASP Conf. Ser. 324: Debris Disks and the Formation of
  Planets, ed. L.~{Caroff}, L.~J. {Moon}, D.~{Backman}, \& E.~{Praton}, 39--+

\bibitem[{{Balbus} \& {Hawley}(1991)}]{Balbus91}
{Balbus}, S.~A. \& {Hawley}, J.~F. 1991, ApJ, 376, 214

\bibitem[{{Bryden} {et~al.}(1999){Bryden}, {Chen}, {Lin}, {Nelson}, \&
  {Papaloizou}}]{Bryden99}
{Bryden}, G., {Chen}, X., {Lin}, D.~N.~C., {Nelson}, R.~P., \& {Papaloizou},
  J.~C.~B. 1999, ApJ, 514, 344

\bibitem[{{Butler} {et~al.}(2006){Butler}, {Wright}, {Marcy}, {Fischer},
  {Vogt}, {Tinney}, {Jones}, {Carter}, {Johnson}, {McCarthy}, \&
  {Penny}}]{Butler06}
{Butler}, R.~P., {Wright}, J.~T., {Marcy}, G.~W., {Fischer}, D.~A., {Vogt},
  S.~S., {Tinney}, C.~G., {Jones}, H.~R.~A., {Carter}, B.~D., {Johnson}, J.~A.,
  {McCarthy}, C., \& {Penny}, A.~J. 2006, ApJ, 646, 505

\bibitem[{{Chiang} {et~al.}(2001){Chiang}, {Joung}, {Creech-Eakman}, {Qi},
  {Kessler}, {Blake}, \& {van Dishoeck}}]{Chiang01}
{Chiang}, E.~I., {Joung}, M.~K., {Creech-Eakman}, M.~J., {Qi}, C., {Kessler},
  J.~E., {Blake}, G.~A., \& {van Dishoeck}, E.~F. 2001, ApJ, 547, 1077

\bibitem[{{Duncan} {et~al.}(1998){Duncan}, {Levison}, \& {Lee}}]{Duncan98}
{Duncan}, M.~J., {Levison}, H.~F., \& {Lee}, M.~H. 1998, AJ, 116, 2067

\bibitem[{{Fleming} \& {Stone}(2003)}]{Fleming03}
{Fleming}, T. \& {Stone}, J.~M. 2003, ApJ, 585, 908

\bibitem[{{Fromang} {et~al.}(2002){Fromang}, {Terquem}, \&
  {Balbus}}]{Fromang02}
{Fromang}, S., {Terquem}, C., \& {Balbus}, S.~A. 2002, MNRAS, 329, 18

\bibitem[{{Gammie}(1996)}]{Gammie96}
{Gammie}, C.~F. 1996, ApJ, 457, 355

\bibitem[{{Glassgold} {et~al.}(1997){Glassgold}, {Najita}, \&
  {Igea}}]{Glassgold97}
{Glassgold}, A.~E., {Najita}, J., \& {Igea}, J. 1997, ApJ, 480, 344

\bibitem[{{Goldreich} \& {Sari}(2003)}]{Goldreich03}
{Goldreich}, P. \& {Sari}, R. 2003, ApJ, 585, 1024

\bibitem[{{Goldreich} \& {Tremaine}(1979)}]{Goldreich79}
{Goldreich}, P. \& {Tremaine}, S. 1979, ApJ, 233, 857

\bibitem[{{Goldreich} \& {Tremaine}(1980)}]{Goldreich80}
---. 1980, ApJ, 241, 425

\bibitem[{{Johnson} {et~al.}(2006){Johnson}, {Goodman}, \&
  {Menou}}]{Johnson06b}
{Johnson}, E.~T., {Goodman}, J., \& {Menou}, K. 2006, ApJ, 647, 1413

\bibitem[{{Laughlin} {et~al.}(2004){Laughlin}, {Steinacker}, \&
  {Adams}}]{Laughlin04}
{Laughlin}, G., {Steinacker}, A., \& {Adams}, F.~C. 2004, ApJ, 608, 489

\bibitem[{{Lecar} {et~al.}(2006){Lecar}, {Podolak}, {Sasselov}, \&
  {Chiang}}]{Lecar06}
{Lecar}, M., {Podolak}, M., {Sasselov}, D., \& {Chiang}, E. 2006, ApJ, 640,
  1115

\bibitem[{{Lin} {et~al.}(1996){Lin}, {Bodenheimer}, \& {Richardson}}]{Lin96}
{Lin}, D.~N.~C., {Bodenheimer}, P., \& {Richardson}, D.~C. 1996, Nature, 380,
  606

\bibitem[{{Lin} \& {Papaloizou}(1993)}]{Lin93}
{Lin}, D.~N.~C. \& {Papaloizou}, J.~C.~B. 1993, {Protostars and Planets III}
  (The University of Arizona Press)

\bibitem[{{Lineweaver} \& {Grether}(2003)}]{Lineweaver03}
{Lineweaver}, C.~H. \& {Grether}, D. 2003, ApJ, 598, 1350

\bibitem[{{Marcy} \& {Butler}(1996)}]{Marcy96}
{Marcy}, G.~W. \& {Butler}, R.~P. 1996, ApJL, 464, L147+

\bibitem[{{Masset}(2001)}]{Masset01}
{Masset}, F.~S. 2001, ApJ, 558, 453

\bibitem[{{Masset} {et~al.}(2006){Masset}, {Morbidelli}, {Crida}, \&
  {Ferreira}}]{Masset06}
{Masset}, F.~S., {Morbidelli}, A., {Crida}, A., \& {Ferreira}, J. 2006, ApJ,
  642, 478

\bibitem[{{Masset} \& {Papaloizou}(2003)}]{Masset03}
{Masset}, F.~S. \& {Papaloizou}, J.~C.~B. 2003, ApJ, 588, 494

\bibitem[{{Matsumura} \& {Pudritz}(2003)}]{Matsumura03}
{Matsumura}, S. \& {Pudritz}, R.~E. 2003, ApJ, 598, 645

\bibitem[{{Matsumura} \& {Pudritz}(2005)}]{Matsumura05}
---. 2005, ApJL, 618, L137

\bibitem[{{Matsumura} \& {Pudritz}(2006)}]{Matsumura06}
---. 2006, MNRAS, 365, 572

\bibitem[{{Mayor} \& {Queloz}(1995)}]{Mayor95}
{Mayor}, M. \& {Queloz}, D. 1995, Nature, 378, 355

\bibitem[{{Menou} \& {Goodman}(2004)}]{Menou04}
{Menou}, K. \& {Goodman}, J. 2004, ApJ, 606, 520

\bibitem[{{Nelson}(2005)}]{Nelson05}
{Nelson}, R.~P. 2005, A\&A, 443, 1067

\bibitem[{{Nelson} \& {Papaloizou}(2004)}]{Nelson04}
{Nelson}, R.~P. \& {Papaloizou}, J.~C.~B. 2004, MNRAS, 350, 849

\bibitem[{{Papaloizou} \& {Larwood}(2000)}]{Papaloizou00}
{Papaloizou}, J.~C.~B. \& {Larwood}, J.~D. 2000, MNRAS, 315, 823

\bibitem[{{Pollack} {et~al.}(1996){Pollack}, {Hubickyj}, {Bodenheimer},
  {Lissauer}, {Podolak}, \& {Greenzweig}}]{Pollack96}
{Pollack}, J.~B., {Hubickyj}, O., {Bodenheimer}, P., {Lissauer}, J.~J.,
  {Podolak}, M., \& {Greenzweig}, Y. 1996, Icarus, 124, 62

\bibitem[{{Pringle}(1981)}]{Pringle81}
{Pringle}, J.~E. 1981, ARA\&A, 19, 137

\bibitem[{{Rafikov}(2002)}]{Rafikov02}
{Rafikov}, R.~R. 2002, ApJ, 572, 566

\bibitem[{{Robberto} {et~al.}(2002){Robberto}, {Beckwith}, \&
  {Panagia}}]{Robberto02}
{Robberto}, M., {Beckwith}, S.~V.~W., \& {Panagia}, N. 2002, ApJ, 578, 897

\bibitem[{{Sano} {et~al.}(2000){Sano}, {Miyama}, {Umebayashi}, \&
  {Nakano}}]{Sano00}
{Sano}, T., {Miyama}, S.~M., {Umebayashi}, T., \& {Nakano}, T. 2000, ApJ, 543,
  486

\bibitem[{{Santos} {et~al.}(2005){Santos}, {Benz}, \& {Mayor}}]{Santos05}
{Santos}, N.~C., {Benz}, W., \& {Mayor}, M. 2005, Science, 310, 251

\bibitem[{{Semenov} {et~al.}(2004){Semenov}, {Wiebe}, \& {Henning}}]{Semenov04}
{Semenov}, D., {Wiebe}, D., \& {Henning}, T. 2004, A\&A, 417, 93

\bibitem[{{Shu} {et~al.}(1994){Shu}, {Najita}, {Ostriker}, {Wilkin}, {Ruden},
  \& {Lizano}}]{Shu94}
{Shu}, F., {Najita}, J., {Ostriker}, E., {Wilkin}, F., {Ruden}, S., \&
  {Lizano}, S. 1994, ApJ, 429, 781

\bibitem[{{Tanaka} {et~al.}(2002){Tanaka}, {Takeuchi}, \& {Ward}}]{Tanaka02}
{Tanaka}, H., {Takeuchi}, T., \& {Ward}, W.~R. 2002, ApJ, 565, 1257

\bibitem[{{Terquem}(2003)}]{Terquem03}
{Terquem}, C.~E.~J.~M.~L.~J. 2003, MNRAS, 341, 1157

\bibitem[{{Thommes}(2005)}]{Thommes05}
{Thommes}, E.~W. 2005, ApJ, 626, 1033

\bibitem[{{Thommes} \& {Murray}(2006)}]{Thommes06}
{Thommes}, E.~W. \& {Murray}, N. 2006, ApJ, 644, 1214

\bibitem[{{Trilling} {et~al.}(1998){Trilling}, {Benz}, {Guillot}, {Lunine},
  {Hubbard}, \& {Burrows}}]{Trilling98}
{Trilling}, D.~E., {Benz}, W., {Guillot}, T., {Lunine}, J.~I., {Hubbard},
  W.~B., \& {Burrows}, A. 1998, ApJ, 500, 428

\bibitem[{{Udry} {et~al.}(2003){Udry}, {Mayor}, \& {Santos}}]{Udry03}
{Udry}, S., {Mayor}, M., \& {Santos}, N.~C. 2003, A\&A, 407, 369

\bibitem[{{Ward}(1992)}]{Ward92}
{Ward}, W.~R. 1992, in Lunar and Planetary Institute Conference Abstracts,
  1491--+

\bibitem[{{Ward}(1997)}]{Ward97}
{Ward}, W.~R. 1997, Icarus, 126, 261

\bibitem[{{Wisdom} \& {Holman}(1991)}]{Wisdom91}
{Wisdom}, J. \& {Holman}, M. 1991, AJ, 102, 1528

\end{thebibliography}
\bibliographystyle{apj}
%
%\clearpage
%
%
%\clearpage
%
%
%\clearpage
%
%
%
%\clearpage
%
%
%
%\clearpage
%
%
%\begin{figure}
%\plotone{figs_chap5/fig8.eps}
%%\includegraphics[width=8cm]{fig8.eps}
%\caption[Gap-opening mass of planets]{The gap-opening mass of
%planets in a typical protostellar disk.  The gap-opening mass is
%about the mass of terrestrial and ice giant planets inside a dead
%zone (\(r<13\) AU), and about a Jupiter mass or larger outside it
%(thick black line). For a planet to open a gap before it hits the
%central star, the gap-opening time has to be shorter than the type I
%migration time (lower solid line) and the viscous evolution time of
%the disk (upper solid line).  To satisfy both conditions, the
%planets have to have a mass larger than the upper solid line
%(shadowed area).  Also plotted are 10 Earth masses and 1 Earth mass
%(dashed lines).  From these, we can see a 10 Earth mass planet can
%open a gap around \(\sim 3.5\) AU inside a dead zone. \label{fig8}}
%\end{figure}
%
%
%\clearpage
%
%
%\clearpage
%
%
%\clearpage
%
\end{document}